\documentclass[journal=jacsat,manuscript=article]{achemso}

\usepackage{chemformula} 
\usepackage[T1]{fontenc} 
\usepackage{lineno,hyperref}
\graphicspath{ {./im/} }
\usepackage{subcaption}
\usepackage{bm,amsmath,amssymb}

\newcommand{\pp}[1]{\left(#1\right)}
\newcommand{\norme}[1]{\left|#1\right|}

\newcommand{\dd}[2]{\frac{\partial #1}{\partial #2}}

{}

\author{Toni El Geitani$^a$}
\author{Shahab Golshan$^a$}
\author{Bruno Blais$^{a,}$}
\email{bruno.blais@polymtl.ca}
\affiliation[Polytéchnique Montréal]
{$^a$Research Unit for Industrial Flows Processes (URPEI), Department of Chemical
Engineering, École Polytechique de Montréal, PO Box 6079, Stn Centre-Ville, Montréal,
QC, Canada, H3C 3A7}

\title[]
  {Towards High-Order CFD-DEM: Development and Validation}

\abbreviations{}

\begin{document}


\begin{abstract}
CFD-DEM is used to simulate solid-fluid systems. DEM models the motion of discrete particles while CFD models the fluid phase. Coupling both necessitates the calculation of the void fraction and the solid-fluid forces resulting in a computationally expensive method. Additionally, evaluating volume-averaged quantities locally restricts particle to cell size ratios limiting the accuracy of the CFD. To mitigate these limitations, we develop a monolithic finite element CFD-DEM solver which supports dynamically load-balanced parallelization. This allows for more stable, accurate and time efficient simulations as load balancing ensures the even distribution of workloads among processors; thus, exploiting available resources efficiently. Our solver also supports high order schemes; thus, allowing the use of larger elements enhancing the validity and stability of the void fraction schemes while achieving better accuracy. We verify and validate our CFD-DEM solver with a large array of test cases: the Rayleigh Taylor instability, particle sedimentation, a fluidized bed, and a spouted bed.
\end{abstract}


\section{Introduction}
Multi-phase flows are prevalent in several industries such as food processing, oil and gas, pharmaceutical industries and many more \cite{ch4,norouzi2016}. Understanding the underlying phenomena that control the behavior of such flows is important to develop state of the art equipment for solid-fluid contactors. Solid-fluid flows and particularly solid-gas flows, are characterized based on the number density of the particles. When this number is small, the flow is termed dilute solid-gas flow and the gas dominates the behavior of the flow. In this case, the solid particles exhibit negligible effects on the gas. When the number density is large, the flow is termed dense solid-gas flow. In this case, the movement of the solid phase becomes mainly controlled by particle-particle collisions. When this number density is bounded and thus lies in between the two previously mentioned cases, the flow is referred to as dispersed flow where the solids constitute the dispersed phase and the fluid constitute the continuous phase. In this case, the flow is affected equally by both phases \cite{ch1}. 

We encounter solid-gas flows in several technological applications such as pollutant control systems, combustion systems, and drying systems. Fluidized and spouted beds are among the common industrial applications which deal with solid-gas flows. Through the application of gas flow at the bottom of the bed, and when the gas flow is large enough to exert sufficient drag on the granular material allowing it to overcome gravity, particles fluidize. In the fluidized state, particles behave as a mixer allowing for a higher mass and heat transfer rates between the two phases resulting in a uniform temperature distribution \cite{vanderHoefM.A2006MMoG}. This results in improved yields and efficiency in chemical and physical processes which employ such beds. 

Fluidized and spouted beds differ in the way they operate. In fluidized beds, the fluid is introduced usually at the entire surface of the inlet while in spouted beds, the fluid enters through a small orifice of the base of the bed. This results in different fluidization behaviors. Contrary to a single fluidizing section in a fluidized bed, spouted beds can be divided into three sections. The spout which is the central core through which the fluid flows. The fountain, which contains the particles entrained by the spout, appears above the bed surface. The annulus is the surrounding annular region of the bed where the particles travel down from the fountain \cite{SahooPranati2013FaSo}.

These beds are complex in their design, building, and operation when compared to other types of beds such as packed beds and stirred tank reactors and are more difficult to scale-up \cite{CoccoRay2014ItF}. However, their advantages for the chemical industries make it vital to predict their behavior in order to accomplish design, scale-up, optimization, and troubleshooting of the processes involved \cite{norouzi2016}. Even though there exist several intrusive and non intrusive experimental techniques to investigate fluidized and spouted beds, their implementation is generally expensive and often infeasible for complex applications. Furthermore, the information that can be extracted from experiments is limited especially at the particle scale. Consequently, modelling can be used to understand the various phenomena occurring in different processes, enabling sensitivity analysis on different input parameters, and testing various configurations and operational conditions at a much cheaper cost \cite{norouzi2016}.

Among the various modeling approaches, unresolved CFD-DEM is often used to study solid-fluid systems. It is an Eulerian-Lagrangian approach in which the flow field is divided into cells larger than the solid particles' size but smaller than the flow field itself. Few commercial and open source CFD-DEM software exist to simulate solid-fluid flows. Usually, CFD and DEM software exist separately and are coupled through a coupling interface. Commercial DEM software include EDEM \cite{edem} and Rocky \cite{rocky} while open source DEM software include LIGGGHTS \cite{liggghts} and Yade \cite{yade}. Commercial CFD software include COMSOL \cite{comsol}, Ansys Fluent \cite{ansys}, StarCCM+ \cite{starccm} among others while open source CFD software include MFiX \cite{mfix} and OpenFOAM \cite{openfoam}. To the best of our knowledge, MFiX, StarCCM+, and PFC \cite{pfc} are the only monolithic CFD-DEM software. However, such software do not support high order elements or dynamic load balancing. 

Unresolved CFD-DEM simulations are computationally expensive and possess some major limitations. The time step of integration is limiting as the DEM scheme is usually explicit and requires the particle time step to be very low (between $10^{-7}$ $s$ to $10^{-4}$ $s$, depending on the size and the stiffness of the particles). Another limitation is the size of the mesh  cells which should usually be at least three times the particle diameter in CFD-DEM simulations \cite{norouzi2016}. This is particularly important to ensure that certain void fraction calculation schemes are valid. The importance of parallelization becomes evident where CFD-DEM can be run on multiple processors to achieve better computational efficiency. Since Lagrangian loads are often distributed unevenly among processors due to the  irregular distribution of particles in fluid cells, computational resources are not fully exploited in an efficient way. Adding load balancing to the recipe allows us to benefit more from parallelization where the work loads are divided evenly among processors \cite{GolshanShahab2021L:Ao}. Currently, there is no CFD-DEM software that implements load balancing in its parallelization strategy as it is complicated and involves balancing the work loads on two interfaces: the CFD and the DEM interfaces. Additionally, more accuracy is required sometimes on the CFD side which is generally, at best, second-order accurate. Therefore, there is a trade off between accuracy and stability in such situation. As such, we propose high order schemes as an important factor in alleviating these challenges. They allow for better accuracy without decreasing the mesh size. Currently existing CFD-DEM software do not exploit high order schemes as they are more complicated and harder to implement than low order methods, and their computational memory requirement is much greater. 

In this paper, we build a high-order FEM stabilized CFD-DEM solver within the open-source software Lethe \cite{BlaisBruno2020LAop} available at \url{https://github.com/lethe-cfd/lethe} which is based on the Deal.II framework \cite{dealII93}. This was achieved by coupling the Volume Averaged Navier-Stokes (VANS) solver \cite{geitani2022high}
with the DEM model of Lethe \cite{GolshanShahab2021L:Ao}. The coupling is built in Lethe and does not require a separate coupling interface. The coupled solver includes all the advantages of the VANS solver \cite{geitani2022high}. It is a fully implicit solver that requires no stability condition on the fluid side. Also, it is globally mass conservative and it accounts for the time variation of the void fraction in the continuity equation. Moreover, it supports both models A and B of the VANS equations. It enables fundamental comparison between the two models in CFD-DEM studies as results obtained from both models are analysed. Furthermore, and to the best of our knowledge, it represents one of the first unresolved CFD-DEM solvers with load balancing capabilities. This enables it to leverage available computing resources by assigning more processors to tasks with greater work loads leading to more efficient utilization of resources. Finally, it is the first high order CFD-DEM solver. As such, coarser meshes can be used without sacrificing accuracy on the CFD side.

The coupling is achieved through the void fraction as well as the particle-fluid forces. In order to verify and validate our unresolved CFD-DEM solver, we consider four test cases. We simulate a particle sedimentation case to determine the particle's terminal velocity. We model a fluidized bed and study the pressure drop and the bed height before and after fluidization. We simulate a heavy fluid on top of a lighter fluid to obtain the Rayleigh Taylor instability and determine its mixing width. Finally, we study a spouted bed and determine the time averaged particle velocity. The obtained results verify correct implementation of the solid-fluid forces, demonstrate global mass conservation, and show physically correct predictions of the hydrodynamics of the systems being studied.

\section{Governing Equations and Coupling Strategy in CFD-DEM}
In the CFD-DEM approach, the incompressible Newtonian fluid phase is modeled as a continuum. It is described by the volume averaged Navier Stokes (VANS) equation which takes into account the dependency of the fluid’s volume within a cell on the solid phase’s volume occupying it. In the present work, these equations are solved using the finite element method in an Eulerian description of matter. On the other hand, the dispersed solid particles are modeled as a discrete phase. The latter is described using the discrete element method \cite{GolshanShahab2021L:Ao}. Therefore, the CFD-DEM model represents an Eulerian-Lagrangian approach for solving multiphase flows.

\subsection{Description of Fluid Phase}
We model the fluid flow using either forms A or B of the volume averaged Navier Stokes equations. The continuity equation for an incompressible fluid is:
\begin{align}
    \rho_f\dd{\pp{\epsilon_f}}{t} + \rho_f\nabla \cdot \pp{\epsilon_f \bm{u}_f} = \mathit{m'} \label{eq::vans_1}
\end{align}
where $\rho_f$ the fluid density, $\bm{u}_f$ is the fluid velocity, $\epsilon_f$ the fluid's void fraction, and $\mathit{m'}$ the volumetric source of mass \cite{GIDASPOW19941}. $\mathit{m'} = 0$ in the case of non reactive flows or when there is no source term added to the simulation.
The momentum equation for model A of the incompressible VANS equations is:
\begin{align}
\rho_f \pp{\dd{\pp{\epsilon_f \bm{u}_f}}{t} + \nabla \cdot \pp{\epsilon_f \bm{u}_f \otimes \bm{u}_f}} = -\epsilon_f \nabla p + \epsilon_f \nabla \cdot \pp{\bm{\tau}_f} + \bm{F_{pf}^A} + \rho_f\epsilon_f \bm{f} 
\label{eq::vans_21}
\end{align}
while that of Model B is:
\begin{align}
\rho_f \pp{\dd{\pp{\epsilon_f \bm{u}_f}}{t} + \nabla \cdot \pp{\epsilon_f \bm{u}_f \otimes \bm{u}_f}} = -\nabla p + \nabla \cdot \pp{\bm{\tau}_f} + \bm{F_{pf}^B} + \rho_f\epsilon_f \bm{f} 
\label{eq::vans_2}
\end{align}
with 
\begin{align}
\bm{\tau}_f = \mu_f \pp{\pp{\nabla \bm{u}_f} + \pp{\nabla \bm{u}_f}^T} - \frac{2}{3} \mu_f \pp{\nabla \cdot \bm{u}_f} \bm{I}\label{eq::stress}
\end{align}
where $p$ is  the pressure, $\bm{\tau}_f$ the deviatoric stress tensor, $\bm{f}$ the external force (ex. gravity), and $\bm{F}_{pf}$ is the momentum transfer term between the solid and fluid phases and includes forces such as drag, virtual mass, Basset force, Saffman lift, and Magnus lift \cite{BlaisBruno2015Otuo}. $\bm{I}$ is the identity or unit tensor.

The difference between the two models lies in the formulation of the momentum equation specifically in the derivation of the pressure and the stress tensor. Additionally, the solid-fluid interactions in model B are explicitly defined whereas, they are implicit in model A \cite{ZHOUZ.Y2010Dpso}. The complete description and derivation of the weak form of the VANS equations using the finite element method (FEM) along with the various stabilization techniques implemented is presented by El Geitani et al. \cite{geitani2022high}. 
All stabilization methods applied support both implicit stabilization where we use the velocity of the current time step in the stabilization formulation and explicit stabilization where we use the velocity at the previous time step.
The last term in Eqs. (\ref{eq::vans_21}) and (\ref{eq::vans_2}) is not added in our solver as we don't solve for gravity in the fluid phase. The consequence of this is the addition of the buoyancy force on the solid phase and the redefinition of the pressure to include the hydrostatic pressure. It is important to note that this force is not explicitly added to the fluid phase of the VANS equations as it is implicit in the pressure. 

The $\gamma$ weight factor parameter for the grad-div stabilization implemented in the VANS equations \cite{geitani2022high} is calculated according to \cite{OLSHANSKII20093975}:
\begin{align}
    \gamma = \nu + c^*\bm{u}_{f_{\Omega_e}}
\end{align}
where $c^*$ depends on the pressure behavior in element $\Omega_e$. Since this information is rarely available, Olshanskii et al. \cite{OLSHANSKII20093975} set $c^*$ to a global constant of order 1. In our solver, we allow $c^*$ to be a user defined value. The choice of $c*$ remains arbitrary in the literature and there is no clear consensus to the optimal value that should be used. Grad-div stabilization plays an important role not only in mass conservation but in the condition number of the linear systems arising when solving the VANS equations. 

In the literature, the derivative of the void fraction with respect to time in the continuity equation is usually set to zero as it introduces major instabilities to the pressure field. In our solver, we found that this quantity can be added to the equation if the density of the fluid is low (eg. for a gas). However, if the fluid is dense (eg. a liquid), this quantity causes the solution to diverge. 

We calculate the void fraction in a cell by:
\begin{align}
\epsilon_f = \frac{v_{element} - v_{particles}}{v_{element}}
\label{eq::void_fraction}
\end{align}
where $v_{element}$ is the volume of an element and $v_{particles}$ is the volume of all particles present within the element. The latter is calculated by the particle centroid method (PCM) due to its simplicity \cite{GolshanShahab2020Raio}. The full description of the void fraction scheme used is explained by El Geitani et al. \cite{geitani2022high}.

\subsection{Description of Solid Phase} 
Newton's second law describes the motion of the solid particles. There are two distinct particle motions in DEM simulations. The particles' translational movement is described by:
\begin{align}
m_{i}\dd{\bm{u}_{p,i}}{t} = \sum_{j \in C_i}\bm{f}_{ij}+\sum_{w}\bm{f}_{w,i} + m_i\bm{g} + \bm{f_{fp,i}}
\label{newton_law}
\end{align}
where $m_i$ is the mass of particle $i$, $\bm{u}_{p,i}$ is the particle $i$'s velocity, $\bm{f}_{ij}$ are the interactions between particle $i$ and particle $j$ and $C_i$ includes all particles in the contact list of particle $i$, $\bm{f}_{w,i}$ are the particle i and walls interactions and $w$ loops over all walls in contact with particle $i$, $\bm{g}$ is the gravity and $\bm{f_{fp,i}}$ are the fluid and particle i interactions which represent the coupling forces between the two phases.
The particles' rotational movement is described by:
\begin{align}
I_i\dd{\bm{\omega}_{p,i}}{t} = \sum_{j \in C_i}\Big(\bm{M}_{ij}^t + \bm{M}_{ij}^r\Big) + \bm{M}_{i}^{ext}
\label{angular_momentum}
\end{align}
where $I_i$ is the moment of inertia of particle $i$, $\bm{\omega}_{p,i}$ is the angular velocity of particle $i$, $\bm{M}_{ij}^t$ and $\bm{M}_{ij}^r$ are respectively the tangential and rolling friction torques due to the contact between particle $i$ and $j$, and $\bm{M}_{i}^{ext}$ denotes all other external torques.

\subsubsection*{Particle-Particle and Particle-Wall Interactions} 
 The soft sphere model is used to calculate contact forces using artificial overlaps and a set of imaginary springs and dashpots and torques resulting from particle collisions \cite{BlaisBruno2019EMiC}. During collision, this overlap as well as the collision force change. The time step for such simulation should be small enough so that the contact between particles is processed in several time intervals \cite{norouzi2016}. We use a non-linear visco-elastic model to calculate the normal and tangential spring and damping constants \cite{GolshanShahab2021L:Ao}. The respective equations are presented in Table \ref{tab:viscoelastic} where $\delta_n$ is the normal overlap, $e$ is the coefficient of restitution, $\nu$ is Poisson's ratio, and i and j represent the particles in contact.

\begin{table}[H]
 \captionof{table}{Normal and tangential spring and damping constants for the visco-elastic model.}
 \label{tab:viscoelastic}
 \centering
\begin{center}
 \begin{tabular}{| c | c |} 
 \hline
\textbf{Parameters} & \textbf{Equations} \\
\hline
Normal spring constant & $k_n = \frac{4}{3}Y_e\sqrt{R_e\delta_n}$
\\
Normal damping coefficient & $\eta_n=-2\sqrt{\frac{5}{6}}\beta\sqrt{S_n m_e}$
\\
Tangential spring constant & $k_t =8G_e\sqrt{R_e\delta_n}$
\\
Tangential damping coefficient & $\eta_t=-2\sqrt{\frac{5}{6}}\beta\sqrt{S_t m_e}$
\\
Effective mass & $\frac{1}{m_e}=\frac{1}{m_i}+\frac{1}{m_j}$
\\
Effective radius & $\frac{1}{R_e}=\frac{1}{R_i}+\frac{1}{R_j}$
\\
Effective shear modulus & $\frac{1}{G_e}=\frac{2(2-\nu_i)(1+\nu_i)}{Y_i}+\frac{2(2-\nu_j)(1+\nu_j)}{Y_j}$
\\
Effective Young's modulus & $\frac{1}{Y_e}=\frac{(1-\nu_i^2)}{Y_i}+\frac{(1-\nu_j^2)}{Y_j}$
\\
Effective Young's modulus & $\frac{1}{Y_e}=\frac{(1-\nu_i^2)}{Y_i}+\frac{(1-\nu_j^2)}{Y_j}$
\\
$\beta$ & $\beta=\frac{\ln e}{\sqrt{\ln^2 e + \pi^2}}$
\\
$S_n$ & $S_n = 2Y_e\sqrt{R_e\delta_n}$
\\
$S_t$ & $S_t = 8G_e\sqrt{R_e\delta_n}$ 
\\
\hline
\end{tabular}
\end{center}
\end{table}

The contact force between two particles is calculated as a combination of normal and tangential contributions\cite{GolshanShahab2021L:Ao}:
\begin{align}
\bm{f}_{ij}=
    \begin{cases}
    \bm{f}_{ij}^n=-(k_n\delta_n)\bm{n}_{ij}-(\eta_n\bm{v}_{rn})\\
     \bm{f}_{ij}^t=-(k_t\delta_t)-(\eta_t\bm{v}_{rt})
    \end{cases}
\end{align}
where $\bm{v}_{rn}$ and $\bm{v}_{rt}$ are the normal and tangential components of the relative contact velocity. For additional details about DEM, we refer the reader to the Lethe-DEM paper \cite{GolshanShahab2021L:Ao}.

In our integration of Newton's equation of motion, we implement the Velocity Verlet integration scheme \cite{DELACROIX2020146}. For more detailed description and explanation of the DEM equations as well as the available integration schemes, we refer the reader to the article by Delacroix et al. \cite{DELACROIX2020146}.

\subsection{CFD-DEM Coupling}
 CFD and DEM are mainly coupled through the void fraction calculation and the fluid-particle interactions. In our code, the CFD and DEM both use the same mesh. This prevents the need to transfer solutions between meshes or localize the particle onto the CFD mesh and significantly reduces the computational cost. Additionally, our solver is fully parallelized. The parallelization of our solver is explained in detail in the Lethe \cite{BlaisBruno2020LAop} and Lethe-DEM \cite{GolshanShahab2021L:Ao} papers for the CFD and DEM components respecively. This parallelization supports adaptive mesh refinement, is flexible and scalable thanks to the method presented by Gassmöller et al. \cite{GassmollerRene2018FaSP} and implemented in the Deal.II library \cite{dealII93}. Every fluid-particle interaction is  calculated using values interpolated at the particle's location  by the FEM interpolation. The summation of individual fluid-particle forces over all particles in a cell results in the force applied on the fluid due to the particles. All forces are calculated only once per time step. The detailed algorithm is shown in the following section.
 
\subsubsection{Coupling Scheme} 
\begin{figure}[ht]
        \centering
        \includegraphics[width=0.9\textwidth]{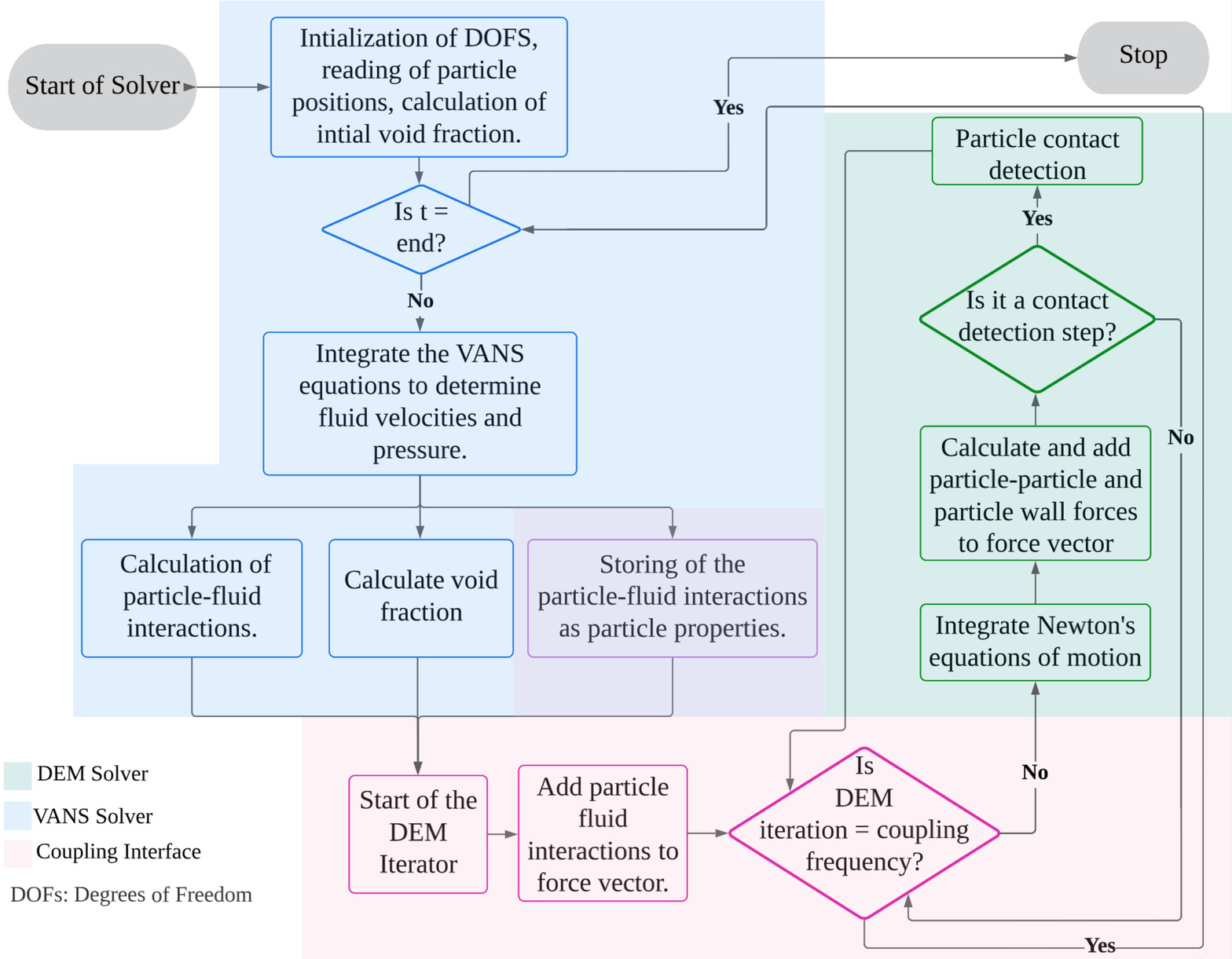}
        \caption{The CFD-DEM coupling scheme.}
        \label{fig:scheme}
\end{figure}

The coupling strategy is given in detail in Fig. \ref{fig:scheme}. It couples the VANS solver \cite{geitani2022high} with Lethe-DEM \cite{GolshanShahab2021L:Ao}. The CFD and DEM solvers are coupled by introducing the DEM iterator within the CFD iterator. Initially, the void fraction is calculated, the VANS equations are solved for the fluid, the fluid-particle interactions are calculated and then the DEM iterator is called to solve Newton's second equation of motion. In order to ensure temporal agreement between the CFD and DEM iterator, we restrict the choice of the time step to the CFD solver. The DEM time step is chosen implicitly by specifying a DEM frequency given as:
\begin{align}
    f_{coupling}=\frac{\Delta t_{CFD}}{\Delta t_{DEM}}
\end{align}
where $f_{coupling}$ is the frequency at which the coupling occurs and $\Delta t$ is the time step. As shown in the VANS article \cite{geitani2022high},
our solver can achieve high order accuracy in both time and space. We can achieve second order or third-order accuracy in time using backward difference formulation (BDF2 or BDF3) and third order accuracy in space using Q2-Q1 finite elements for velocity and pressure respectively.
On the CFD side, our code is unconditionally stable since we use an implicit formulation in our evaluation of the VANS solution. On the DEM side, and in order to produce valid results, the DEM time step should be small enough to accurately capture particle-particle interactions. Since the fluid modeling is based on evaluating locally volume-averaged quantities, an element $(\Omega_e)$ volume should be at least one order of magnitude greater than the particle volume $(V_{\Omega_e}> 10V_p)$ \cite{PEPIOT2012104}.

For the inter-phase coupling, and as we aim to solve dense solid-fluid systems, we assume 4 way coupling between the solid and fluid phases. In this case, the fluid affects the particles' motion and the particles affect the fluid flow. In addition to this, particles motion is also affected by collisions between particles and between particles and walls where these interactions play an important role in characterizing the particles' behavior \cite{Elghobashi}. 

\subsubsection*{Load Balancing}
Our CFD-DEM coupled solver supports load balancing. Load balancing allows the distribution of all tasks among all available processors such that the overall computational time is optimized.  Load balancing was possible thanks to the p4est \cite{p4est} and Deal.II \cite{dealII93} finite element library. P4est allows for a user-specified weight function that returns a non-negative integer weight for each finite element. Thus, instead of partitioning the work into a uniform number of elements, it distributes it evenly by weight. This is important as sometimes elements have varying mathematical unknowns to store and compute on \cite{p4est}. This is true for CFD-DEM as the number of particles is not equal among elements. In the CFD-DEM solver of Lethe, each cell is assigned a weight based on its computational load. The sum of these weights is then distributed evenly among the available processors. In the absence of particles, the default cell weight is 1000. In CFD-DEM, we attribute a weight for a particle. The overall cell weight becomes the sum of the cell weight and all particles' weights in the cell. Consequently, a cell with more particles will normally have a larger weight.

The application of load balancing in CFD-DEM software is relatively new. To our knowledge, our CFD-DEM solver demonstrates one of the first solvers with load balancing capabilities that take into account both physics. Further studies must be performed in order to determine the optimum particle to cell weight ratios as this is affected by several factors. For example, a simulation where particle-particle contact forces are frequent usually requires a larger particle weight, whereas the same simulation with Q2-Q1 elements instead of Q1-Q1 elements will require a larger cell weight.

\subsubsection{Particle-fluid Interactions} 
 There exists several important forces that should be accounted for in the coupling process. These include the drag force, the buoyancy force, the pressure force, and the shear force. Since we are mainly interested in gas-solid flows in this work, we neglect virtual mass force, Basset force and lift force \cite{GolshanShahab2020Raio}.
 
\subsubsection*{Drag Force} 
It is the force experienced due to the relative motion of the particle and fluid around it. We determine the drag force based on:
\begin{align}
    \bm{F_{D}} = \beta(\bm{u}_f-\bm{u}_{p,i})
    \label{eq::drag}
\end{align}
where $\bm{u}_i$ is the particle i's velocity, and $\beta$ is the inter-phase momentum exchange coefficient. According to Gidaspow, the interphase momentum exchange coefficient $\beta$ differs between models A and B \cite{GIDASPOW19941}:
\begin{align}
    \beta_{B} = \frac{\beta_A}{\epsilon_f}
    \label{eq::beta}
\end{align}

On the other hand, Zhou et al. \cite{ZHOUZ.Y2010Dpso} implement the same drag force for both models. In our derivation, we use the same drag force for both models. This is because only when considering the same drag force are the models A and B mathematically equivalent. When simulating a stationary solid phase, as the case of the packed bed which we simulated \cite{geitani2022high},
we use the assumption of Eq. (\ref{eq::beta}) as we only consider drag and buoyancy. Eqs. (\ref{eq::fb}) and (\ref{eq::fa}) demonstrate that the pressure gradient force should be applied back to the fluid phase only in model B. As such, not implementing this important force in the packed bed for model B necessitates the use of the momentum exchange coefficient as written in Eq. (\ref{eq::beta}). This penalizes the pressure in model B by a factor of $\epsilon_f$ in order to compensate for the pressure gradient force resulting in a good pressure drop in the bed. This is not the case for model A as the pressure gradient force is already implicitly applied by the term $\epsilon_f \nabla p$ of the momentum equation.

We calculate the drag force on a single particle as:
\begin{align}
    \bm{F_{D}} = \frac{1}{2}\rho_f C_D A_{ref} \norme{\bm{u}_{f,p}-\bm{u}_{p,i}}(\bm{u}_{f,q}-\bm{u}_{p,avg})
\end{align}
where $\bm{u}_{f,p}$ is the interpolated fluid velocity at the particle's location calculated at the previous time step, $\bm{u}_{f,q} = \bm{u}_f$ is the fluid velocity at the quadrature point, and $\bm{u}_{p,avg}$ is the average particles' velocity in the cell both calculated at the current time step. We use a $Q_n$ interpolation that is homogeneous with the FEM scheme in order to interpolate the velocity at the particle's location. $A_{ref}$ is the particle's reference area which is taken as the cross-section:
\begin{align}
    A_{ref} = \pi r_p^2
\end{align}
where $r_p$ is the particle's radius and  $C_D$ is the drag coefficient. At the time of writing, our code supports Dallavalle \cite{dallavalle}, DiFelice \cite{DiFeliceR1994Tvff}, Rong \cite{RongL.W2013Lsof}, Koch and Hill \cite{JAJCEVIC2013298}, Beetstra \cite{beetstradrag}, and Gidaspow \cite{GIDASPOW19941} drag models.
The drag coefficient for Dallavalle model \cite{dallavalle}:
\begin{align}
    C_D = \Big(0.63 + \frac{4.8}{\sqrt{Re_p}}\Big)^2
\end{align}
and $Re_p$ is the particle Reynolds number and is expressed as:
\begin{align}
    Re_p =\frac{\rho_f \epsilon_f \norme{\bm{u}_f-\bm{u}_{p,i}}d_p}{\mu_f}
\end{align}
where $\mu_f$ is the fluid's dynamic viscosity and $d_p$ is the particle's diameter. The drag coefficient for the Di Felice model \cite{DiFeliceR1994Tvff} is given:
\begin{align}
    C_{D} = \Big(0.63 + \frac{4.8}{\sqrt{Re_p}}\Big)^2 \epsilon_f^{2-\big[3.7 - 0.65 e^{\big(\frac{-(1.5-log_{10}(Re_p))^2}{2}\big)}\big]}
\end{align} while that for the Rong et al. model is given as \cite{RongL.W2013Lsof}:
\begin{align}
    C_{D} = \Big(0.63 + \frac{4.8}{\sqrt{Re_p}}\Big)^2 \epsilon_f^{2-\big[2.65(\epsilon_f + 1) - (5.3-3.5\epsilon_f)\epsilon_f^2 e^{\big(\frac{-(1.5-log_{10}(Re_p))^2}{2}\big)}\big]}
\end{align}
The Koch and Hill drag model is derived based on results obtained from lattice-Boltzmann simulations. It is calculated according to Eq. (\ref{eq::drag}) where the momentum exchange coefficient $\beta$ is determined as \cite{JAJCEVIC2013298}:
\begin{align}
    \beta = \frac{18 \mu_f \epsilon^2_f \epsilon_p}{d^2_p}\Big(F_0(\epsilon_p) + \frac{1}{2}F_3(\epsilon_p) Re_p\Big)\frac{V_p}{\epsilon_p}
\end{align}
where $\epsilon_p = 1- \epsilon_f$ is the particles' void fraction, $V_p$ is the particle's volume, and where 
\begin{align}
    F_0(\epsilon_p) = \begin{cases}
    \frac{1+3\sqrt{\frac{\epsilon_p}{2}}+\frac{135}{64}\epsilon_p ln(\epsilon_p)+16.14\epsilon_p}{1+0.681\epsilon_p-8.48\epsilon_p^2+8.14\epsilon_p^3} & \epsilon_p < 0.4 \\
    \frac{10 \epsilon_p}{\epsilon_f^3} & \epsilon_p \geq 0.4
    \end{cases}
\end{align}
and 
\begin{align}
    F_3(\epsilon_p) = 0.0673 + 0.212\epsilon_p + \frac{0.0232}{\epsilon_f^5}
\end{align}
The Beetstra \cite{beetstradrag} drag model was obtained from lattice-Boltzmann simulations and the drag force for a single particle is expressed as:
\begin{align}
    \bm{F_{D}} = 3 \pi \mu_f d_p \epsilon_f \bm{u}_f F_i
\end{align}
where $F_i$ is the normalized drag for a mono-dispersed system and is defined as:
\begin{align}
    F_i = \frac{10 \epsilon_p}{\epsilon_f^2} + \epsilon_f^2(1+1.5\epsilon_p^{0.5})+\frac{0.413Re}{24\epsilon_f^2}\Bigg[\frac{\epsilon_f^{-1}+3\epsilon_p \epsilon_f +8.4Re_p^{-0.343}}{1+10^{3\epsilon_p}Re_p^{-(1+4\epsilon_p)/2}}\Big]
\end{align}
The Gidaspow \cite{GIDASPOW19941} drag model is a combination between the Ergun equation and Wen-Yu drag model. It is defined as: 
\begin{align}
    \beta = \begin{cases}
    \frac{150(1-\epsilon_f)^2 \mu_f}{\epsilon_f d_p^2} + \frac{1.75(1-\epsilon_f)\rho_f (\bm{u}_f - \bm{u}_{p})}{d_p} & \epsilon_f < 0.8 \\
    \frac{3}{4}C_D\frac{\epsilon_f \norme{\bm{u}_f - \bm{u}_p}\rho_f (1-\epsilon_f)}{d_p}\epsilon_f^{-2.65} & \epsilon_f \geq 0.8
    \end{cases}
\end{align}
where for the case of $\epsilon_f \geq 0.8$, the drag coefficient $C_D$ is determined as:
\begin{align}
C_D = 
    \begin{cases}
    \frac{24}{Re_p}\Big(1 + 0.15(Re_p)^{0.687}\Big) & Re_p < 1000 \\
    0.44 & Re_p \geq 1000
    \end{cases}
\end{align}
For additional information about the different drag models available, we refer the reader to the article by Norouzi et al. \cite{NorouziHamidReza2021Otdf} or by Bérard et al. \cite{BerardAriane2020Emic}.

\subsubsection*{Buoyancy Force}
It is the force exerted by the fluid on the submerged particle that opposes its weight. It is given by the following equation \cite{GIDASPOW19941}:
\begin{align}
    \bm{F}_{B}=-\rho_f V_p \bm{g}  
\end{align}
where $V_p$ is the particle's volume. The buoyancy force becomes important in fluidized and spouted beds when the ratio of particle to fluid densities is significant. 
The buoyancy force is only applied on the particles, as the pressure calculated using the VANS equations includes the hydrostatic pressure. 

\subsubsection*{Undisturbed Flow Forces}
These forces are given by \cite{BerardAriane2020Emic}:
\begin{align}
    \bm{F}_{\nabla p}= -V_p\Big(\dd{p}{x}\Big)=-V_p\nabla P\\
    \bm{F}_{\nabla \cdot \tau}= -V_p\Big(\dd{\tau}{x}\Big)=-V_p\nabla \cdot\tau
\end{align}
where $\bm{F}_{\nabla p}$ and $\bm{F}_{\nabla \cdot \tau}$ are the pressure and shear forces respectively. For Model A, these forces are implicitly added to the fluid since the void fraction multiplies the pressure and stress tensor gradients, and therefore should only be added explicitly to the particles. For Model B, it should be explicitly added for both solid and fluid phases.

All of the fluid-particle interactions are applied to both the solid and fluid phases to ensure that Newton's third law of motion is respected. The expression of the particle-fluid interactions for the VANS equations are  $\bm{F}_{{pf}}$ for the force applied on the fluid phase and $\bm{f}_{fp}$ for the force applied on the solid phase. For model B, these forces become:
 \begin{align}
 \label{eq::fb}
     \bm{F}_{{pf}_{B}} = \frac{1}{\Delta V_{\Omega_e}}\sum_{N_p}\Big(\bm{F}_{D_{B}} + \bm{F}_{\nabla p} + \bm{F}_{\nabla \cdot \tau}\Big) \\
     \bm{f}_{{fp}_{B},i} = \Big(\bm{f}_{D_{B},i} + \bm{f}_{B,i} + \bm{f}_{\nabla p,i} + \bm{f}_{\nabla \cdot \tau,i}\Big)
 \end{align}
where $V_{\Omega_e}$ is the volume of the finite element $\Omega_e$. The particle-fluid force of model A is given by:
 \begin{align}
  \label{eq::fa}
     \bm{F}_{{pf}_{A}} = \frac{1}{\Delta V_{\Omega_e}}\sum_{N_p}\Big(\bm{F}_{D_{A}}\Big)\\
      \bm{f}_{{fp}_{A},i} = \Big(\bm{f}_{D_{A},i}+ \bm{f}_{B,i} + \bm{f}_{\nabla p,i} + \bm{f}_{\nabla \cdot \tau,i}\Big)
 \end{align}

\section{Verification and Validation of the CFD-DEM Model}
In order to verify and validate our model, we simulate several test cases. The series of test cases include particle sedimentation, a fluidized bed, the Rayleigh Taylor instability and a spouted bed.

\subsection{Particle Sedimentation Test Case}
We simulate a dense particle initially at rest falling in a stagnant liquid and we measure the instantaneous velocity of the particle. We then compare the velocity obtained from the simulation with that calculated by numerically solving the following ordinary differential equation (ODE):
\begin{align}
    m_p\frac{\partial u}{\partial t}= \Big( \bm{F}_D - V_p (\rho_p -\rho_f)\bm{g}  \Big)
    \label{eq::ODE}
\end{align}
where $m_p$ is the particle's mass and $\rho_p$ is the particle's density. After some time, the particle should reach a constant velocity known as the settling or terminal velocity. This occurs when the drag balances the gravitational force and buoyancy \cite{GIDASPOW19941}. The settling velocity $\bm{v}_r$ is given as:
\begin{align}
    \bm{v}_r = \sqrt{\frac{V_p (\rho_p-\rho_f)\bm{g}}{\frac{1}{2}\rho_f C_D A_{ref}}}
\end{align}
where $C_D$ is the drag coefficient given based on the drag model used. We use the DiFelice drag model. 
\subsubsection{Simulation Setup}
The cylinder in which the particle will fall should be large enough so that its walls have no effect on the sedimentation of the particle. For this, we choose a cylinder with a radius 25 times greater than the particle's diameter. The particle was inserted in the middle at a height of 0.08 from the center of the cylinder having the following coordinates (0.08,0,0). We chose a very coarse mesh as to avoid velocity fluctuations as the particle moves across cells. Only the drag and buoyancy forces were enabled for this simulation as to respect the ODE of Eq. (\ref{eq::ODE}). The physical properties chosen are typical values often used for water as a fluid and glass beads as the solids. We used Q1-Q1 second order finite elements in velocity and pressure. All simulation parameters are given in Table \ref{tab:setup1}. 
\begin{table}[H]
 \captionof{table}{Physical and numerical parameters for the particle sedimentation test case.}
 \label{tab:setup1}
\centering
\resizebox{\textwidth}{!}{%
 \begin{tabular}{l l l l} 
 \hline
 \textit{Simulation control}\\
 End time ($s$) & 0.5\\
Coupling frequency & $10^{2}$ & CFD time step ($s$) & $10^{-3}$ \\
 \\
 \textit{Geometry}\\
 Bed Height ($mm$) & 100 &  Bed Radius ($mm$) & 50 \\
 Wall Thickness ($mm$) & 0 &Mesh Type & dealii::subdivided-cylinder\\ Mesh Refinement & 1 & Mesh Subdivisions in x & 5 
 \\
 \\
 \textit{Particles}\\
 Number & $1$ & Diameter(m) & $2\times10^{-3}$\\
 Density($kg/m^3$) & 2500 & Young Modulus ($N/m^2$) &  $1\times10^6$\\
 Particle-particle poisson ratio & 0.3 & Particle-wall poisson ratio & 0.3 \\
 Particle-particle restitution coefficient & 0.2 & Particle-wall restitution coefficient & 0.2 \\
 Particle-particle friction coefficient & 0.1 & Particle-wall friction coefficient & 0.1 \\
 Particle-particle rolling friction & 0.2 & Particle-wall rolling friction & 0.3 \\
 \\
 \textit{Gas phase}\\
 Viscosity ($Pa\cdot s$) & $1.005\times10^{-6}$ & Density ($kg/m^3$) & 997 \\
 Inlet Velocity ($m/s$) & 0 & Void fraction smoothing factor $L^2$ & $0$\\
 \\
 \textit{Linear Solver}\\
 Method & GMRES & Max iterations & 5000\\
 Minimum residual & $1\times10^{-11}$ & Relative residual & $10
 \times10^{-3}$\\
 ILU preconditioner fill & 1 & ILU preconditioner absolute tolerance & $1\times10^{-14}$\\
 ILU preconditioner relative tolerance & 1\\
 \\
 \textit{Non-linear solver}\\
 Tolerance & $1\times10^{-9}$ & Max iterations & 10\\
 \hline
 \end{tabular}}
\end{table}

\subsubsection{Results and Discussion}
\begin{figure}[H]
        \centering
        \includegraphics[width=0.8\textwidth]{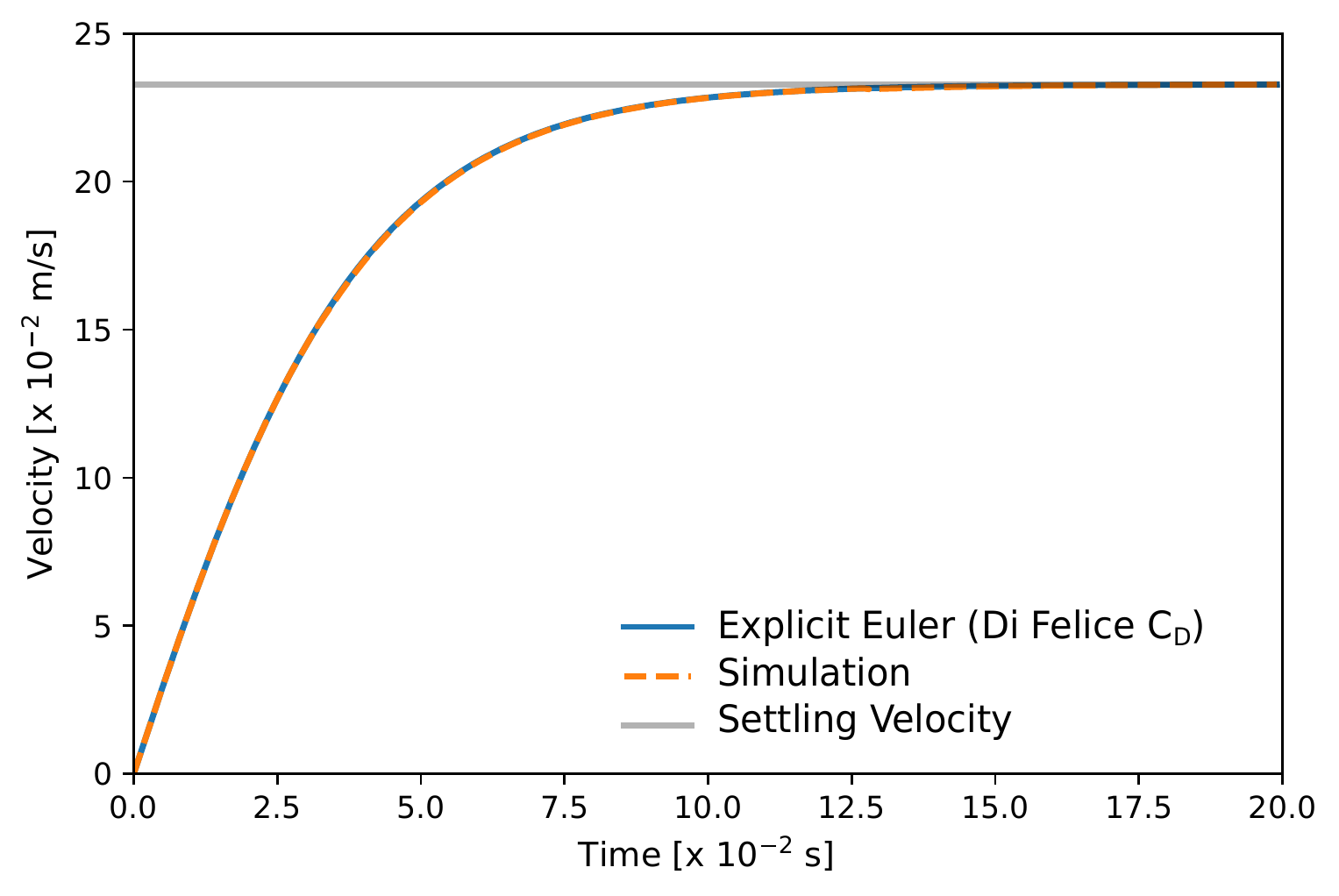}
        \caption{Instantaneous velocity of particle during sedimentation.}
        \label{fig:sedimentation}
\end{figure}

We show the instantaneous velocity of the particle in Fig. \ref{fig:sedimentation}. Based on the obtained results, the velocity profile of the particle follows the analytical solution until it reaches the correct value of the settling velocity $\bm{v}_r = 0.2328 m/s$. 
This simple test case allows the validation of our drag and buoyancy implementation. 

\subsection{Fluidized Bed Test Case}
We study the pressure drop as a function of inlet velocity in a fluidized bed. We compare the measured pressure drop with that obtained from the Ergun equation. The Ergun equation is a correlation that predicts the pressure drop in a packed bed and is given by \cite{Ergun}:
\begin{align}
    \Delta p = \frac{150(1-\epsilon_f)^2 \bm{u}_f \mu_f H_b}{\epsilon_f^3 d_p^2} + \frac{1.75(1-\epsilon_f)\rho_f \bm{u}_f H_b}{\epsilon_f^3 d_p}
    \label{eq::ergun}
\end{align}
where $H_b$ is the bed's height.
After fluidization, the pressure drop becomes constant and proportional to particle weight held by the fluid and is given as:
\begin{align}
    \Delta p =\frac{N_p V_p (\rho_p -\rho_f)g}{A_{b}}
    \label{eq::dpf}
\end{align}
where $N_p$ is the total number of particles in the bed and $A_{b}$ is the bed's cross sectional area. In a fluidized bed, the pressure drop is only the contribution of the drag and buoyancy forces. We simulate the fluidized bed test case using both Q1-Q1 and Q2-Q1 elements to demonstrate the high order capabilities of our CFD-DEM solver. For the discretization in time, we use the first order backward difference scheme (BDF1).

\subsubsection{Simulation Setup}
We first use the DEM solver to fill a cylindrical bed with 200,000 spherical particles with a diameter of 500 $\mu m$. The particles are packed at a distance of 0.04 $m$ above the inlet to allow the fluid flow to develop before entering the packing. This also eliminates any effects the void fraction might have at the inlet. The DEM simulation had a final time of 1 $s$ to allow the particles to settle and for their kinetic energy to dissipate. Air flow is then introduced at the bottom of the bed at varying inlet velocities from 0.02 $m/s$ to 0.28 $m/s$ with increments of 0.02 $m/s$. An additional velocity of 0.15 $m/s$ was added as it represents the value close to the minimum fluidization, thus it allows us to better understand the behavior at minimum fluidization.  We choose a smoothing length for the void fraction equivalent to $4$ $d_p^2$. We study the fluidization curve of the bed obtained using the Di Felice \cite{DiFeliceR1994Tvff} drag model. The simulation parameters are given in Table \ref{tab:setup2}.

\begin{table}[H]
 \captionof{table}{Physical and numerical parameters for the fluidized bed test case.}
 \label{tab:setup2}
\centering
\resizebox{\textwidth}{!}{%
 \begin{tabular}{l l l l} 
 \hline
 \textit{Simulation control}\\
 End time ($s$) & 5 & Time discretization scheme & BDF1\\
Coupling frequency  & $10^{2}$ & CFD time step ($s$) & $10^{-3}$ \\
 \\
 \textit{Geometry}\\
 Bed Height ($mm$) & 400 &  Bed Radius ($mm$) & 10 \\
 Wall Thickness ($mm$) & 0 &Mesh Type & dealii::subdivided-cylinder\\ Mesh Refinement & 2 & Mesh Subdivisions in x & 40 
 \\
 \\
 \textit{Particles}\\
 Number & $2\times10^5$ & Diameter(m) & $5\times10^{-4}$\\
 Density($kg/m^3$) & 1000 & Young Modulus ($N/m^2$) &  $1\times10^6$\\
 Particle-particle poisson ratio & 0.3 & Particle-wall poisson ratio & 0.3 \\
 Particle-particle restitution coefficient & 0.9 & Particle-wall restitution coefficient & 0.9 \\
 Particle-particle friction coefficient & 0.1 & Particle-wall friction coefficient & 0.1 \\
 Particle-particle rolling friction & 0.2 & Particle-wall rolling friction & 0.3 \\
 \\
 \textit{Gas phase}\\
 Viscosity ($Pa\cdot s$) & $1\times10^{-5}$ & Density ($kg/m^3$) & 1 \\
 Inlet Velocity ($m/s$) & [0.02-0.28] & Void fraction smoothing factor $L^2$ & $1.25\times10^{-6}$\\
 \\
 \textit{Linear Solver}\\
 Method & GMRES & Max iterations & 5000\\
 Minimum residual & $10^{-10}$ & Relative residual & $10
 \times10^{-3}$\\
 ILU preconditioner fill & 1 & ILU preconditioner absolute tolerance & $10^{-14}$\\ 
 ILU preconditioner relative tolerance & 1-1.1 & max krylov vectors & 1000 | 2000\\
 \\
 \textit{Non-linear solver}\\
 Tolerance & $1\times10^{-9}$ & Max iterations & 10\\
 \hline
 \end{tabular}}
\end{table}

There exists various correlations to predict the minimum fluidization velocity. Based on the particles’ diameter and density, we compare the simulation results with different correlations. We used Ergun \cite{Ergun}, Wen-Yu \cite{wen1966generalized}, and Noda et al. \cite{NODA1986149} correlations for the minimum fluidization velocity. For a more detailed review on the different available correlations, we refer the reader to the article by Anantharaman et al. \cite{ANANTHARAMAN2018454}. The Ergun minimum fluidization $Re_{mf}$ is defined as:

\begin{align}
    Re_{mf}= \Bigg(\Big(\frac{42.86(1-\epsilon_{mf})}{\phi}\Big)^2+0.571 \epsilon_{mf}^3 \phi Ar \Bigg)^{0.5} - \Bigg(\frac{42.86(1-\epsilon_{mf})}{\phi}\Bigg)
    \label{eq::umf1}
\end{align}

where $\epsilon_{mf}$ is the void fraction at minimum fluidization, $\phi$ is the sphericity of the particles ($\phi = 1$ for spherical particles), and $Ar$ is the Archimedes number and is defined as:
\begin{align}
    Ar = \frac{g \rho_f (\rho_p - \rho_f) d_p^3}{\mu_f^2}
\end{align}

The Wen-Yu minimum fluidization $Re_{mf}$ is defined as:
\begin{align}
    Re_{mf}= \Big(33.7^2 + 0.0408 Ar\Big)^{0.5} - 33.7
    \label{eq::umf2}
\end{align}

The Noda et al. minimum fluidization $Re_{mf}$ is defined as:
\begin{align}
    Re_{mf}= \Big(19.29^2 + 0.0276 Ar\Big)^{0.5} - 19.29
    \label{eq::umf3}
\end{align}

The minimum fluidization velocity $(U_{mf})$ is obtained from $Re_{mf}$ by:
\begin{align}
    Re_{mf} = \frac{\rho_f U_{mf} d_p}{\mu_f}
    \label{eq::re}
\end{align}

For this case, we found the minimum fluidization velocity using Eq. (\ref{eq::umf1}) to be $U_{mf} = 0.137$ $m/s$, Eq. (\ref{eq::umf2}) to be $U_{mf} = 0.135$ $m/s$, and Eq. (\ref{eq::umf3}) to be $U_{mf} = 0.147$ $m/s$. The fluidization pressure drop is calculated using Eq. (\ref{eq::dpf}) to be $\Delta p_{fluidization} = 408.34$ $Pa$. Grad-div stabilization was particularly interesting in this case when solving model B of the VANS equations where stability was dependant on the choice of $c^*$. Model A showed an enhanced stability with an optimal choice of $c^*$ taken to be the size of the element for all cases. This was not the case for model B, where the solver was unstable for some cases. Hence, the choice of $c^* = 1$.

\subsubsection{Results and Discussion}
\begin{figure}[H]
     \centering
     \begin{subfigure}[b]{0.7\textwidth}
         \centering
         \includegraphics[width=\textwidth]{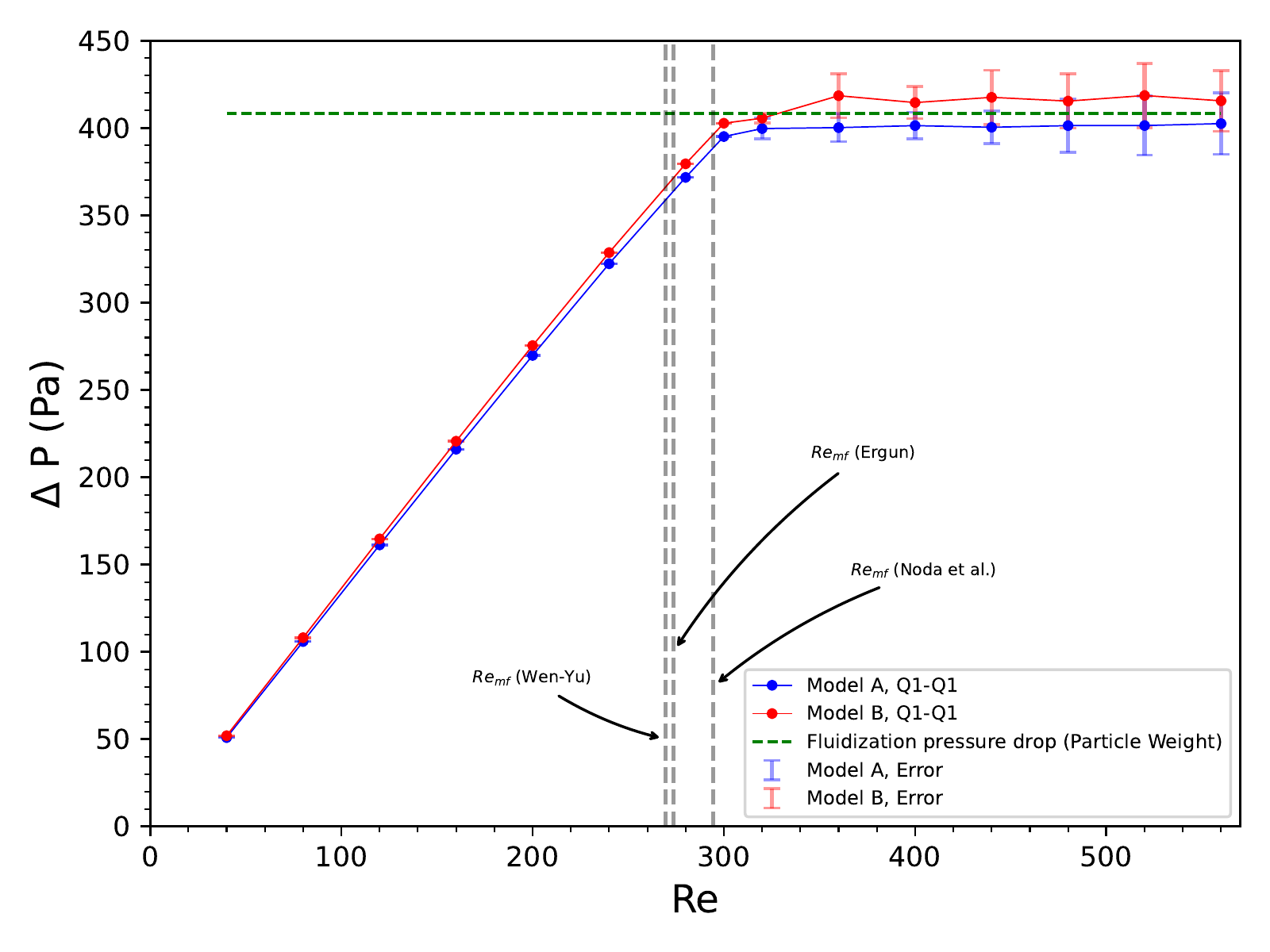}
         \caption{pressure drop using Q1-Q1 elements}
         \label{fig:pressure_drop_Q1}
     \end{subfigure}
     \\
     \begin{subfigure}[b]{0.7\textwidth}
         \centering
         \includegraphics[width=\textwidth]{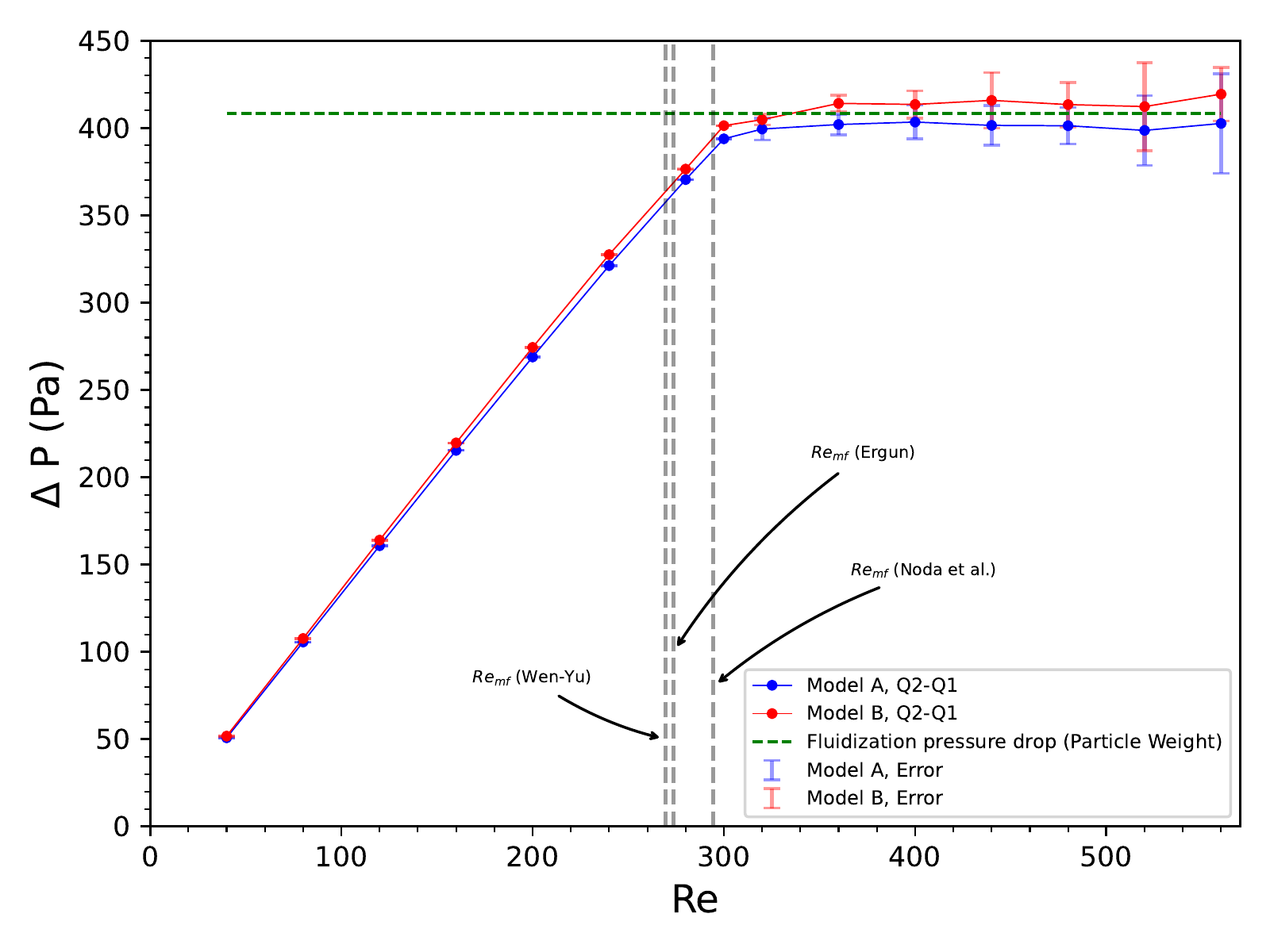}
         \caption{pressure drop using Q2-Q1 elements}
         \label{fig:pressure_drop_Q2}
     \end{subfigure}
        \caption{Pressure drop in fluidized bed as a function of inlet Re number.}
        \label{fig:fluidized_bed_dp}
\end{figure}

We calculate the pressure drop instantaneously, and we average it over 0.5 $s$ after the pressure stabilizes. Fig. \ref{fig:fluidized_bed_dp} shows the pressure drop as a function of the different inlet velocities for both models A and B of the VANS equations for different finite element orders. The convergence of model B with Q2-Q1 elements was difficult as the system of model B is already stiff and using high order elements exacerbates the stiffness of the system. The stiffness of model B comes from explicitly adding the pressure and shear forces to the fluid. For model B with Q2-Q1 elements, few simulations with inlet velocities above minimum fluidization were not converging to the tolerance of $10^{-9}$ for longer simulation times. As a result, some of the Q2-Q1 model B cases were simulated for 3 $s$ instead of 5 $s$. Model A is considerably more stable than model B in CFD-DEM simulations as convergence was obtained easily for the different elements orders. The error bars shown represent the standard deviation over a period of 0.5 $s$. 

For both finite element orders, it can be seen that models A and B result in a relatively close pressure drop in the bed. Model B results in a slightly larger standard deviation and that can be attributed to its instantaneous pressure drop fluctuating to larger magnitudes than that of model A. The fluctuations in pressure drop for both models is slightly higher for second order elements. Since higher order elements capture better the turbulence in the flow, we expect to observe higher fluctuations in the pressure drop.

\begin{figure}[H]
        \centering
        \includegraphics[width=0.7\textwidth]{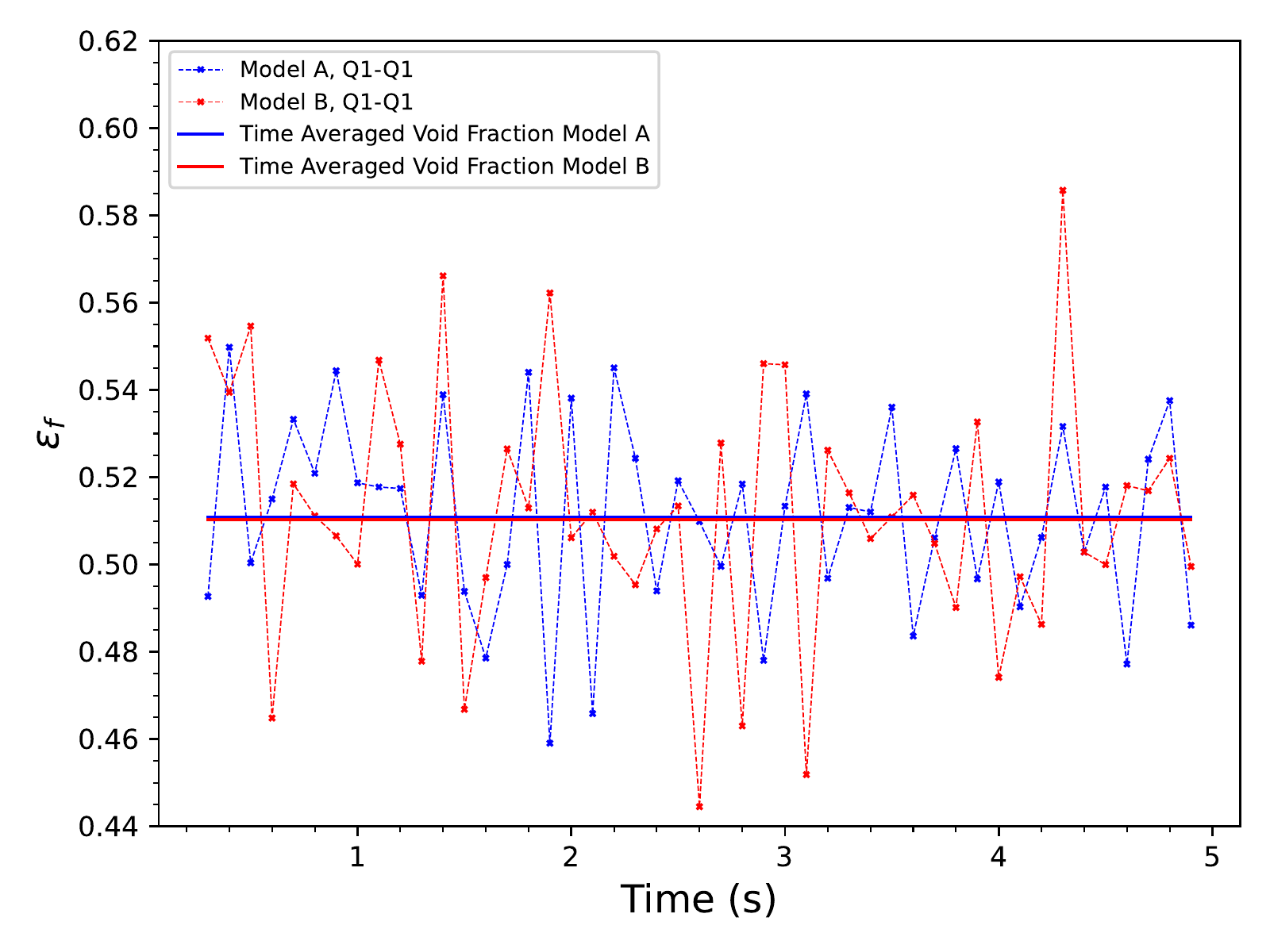}
        \caption{Instantaneous bed void fraction for inlet velocity of 0.28 $m/s$.}
        \label{fig:instant_bed_height}
\end{figure}

Fig. \ref{fig:instant_bed_height} presents the instantaneous bed void fraction for both models using Q1-Q1 elements and an inlet velocity of 0.28 $m/s$. The void fraction of the bed is a representation of its packing height. The void fraction of both models is highly oscillatory explained by the slug flow behavior of the fluidized bed where large bubbles form at the inlet of the bed and break at the outlet of the packing region. Models A and B do not lead to the same instantaneous physical behavior. This is in part due to the chaotic behavior of a fluidized bed. In model A, the shear and pressure gradient forces are calculated and added to the fluid phase according to cell properties such as cell void fraction and cell pressure drop. However, in model B these forces are calculated on individual particles and then lumped up with the drag force before being added to the fluid phase. The time averaged height of both models is almost equal and as such, the beds behavior is statistically similar.

\begin{figure}[H]
        \centering
        \includegraphics[width=0.7\textwidth]{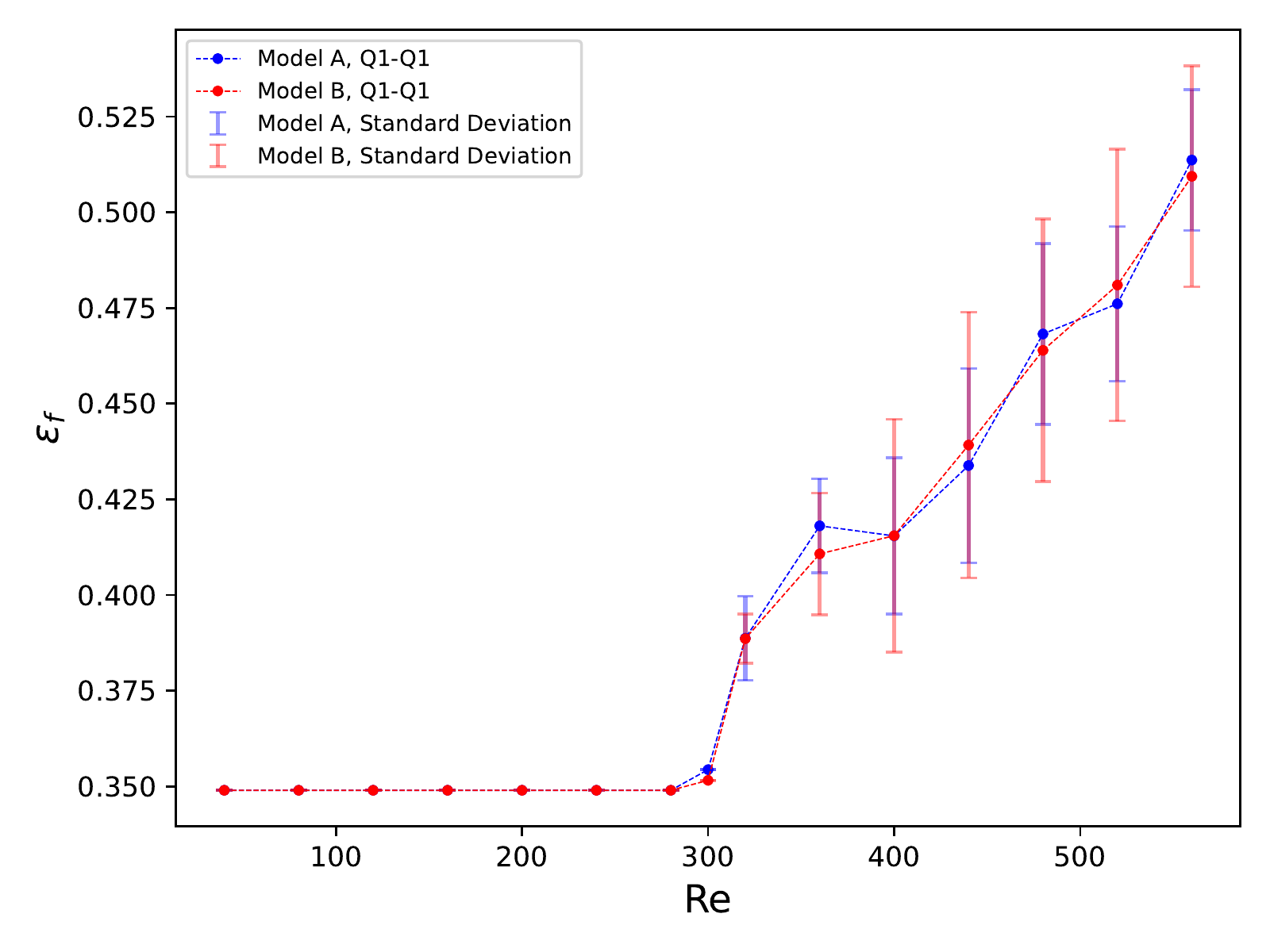}
        \caption{Average bed void fraction as a function of inlet Re number using Q1-Q1 elements.}
        \label{fig:average_bed_height}
\end{figure}

For Q1-Q1 elements, we compare the time averaged void fraction in the bed for both models as shown in Fig. \ref{fig:average_bed_height}. The void fraction was averaged over a period of 1 $s$.

The time averaged bed void fraction is relatively close in both models for the different inlet velocities. The variation at each inlet velocity over 1 $s$ is shown as the standard deviation which represents the magnitude of fluctuation of the bed height over this period. Model B fluctuations are more pronounced when compared to those of model A as demonstrated by the standard deviation in Fig. \ref{fig:average_bed_height}. Therefore, model A and B, though mathematically equivalent are physically different when considering the instantaneous behavior of the fluidized bed. This is caused by the different treatment of the pressure force.

\subsection{The Rayleigh-Taylor Instability Test Case}
We study the evolution of the Rayleigh-Taylor instability as a function of different fluid and particle densities. This is a well known case in which a heavy fluid is located above a light fluid. The difference in densities makes this case an inherently unstable problem. The fluids start mixing together under the effect of gravity. Initially, an exponential instability growth period occurs after which the mixing layer is described according to the following \cite{SniderD.M2001AITM}:
\begin{align}
    h = \alpha A g t^2
    \label{eq::mixing_layer}
\end{align}
where h is the width of the mixing layer, g is the gravitational acceleration, t is the time and A is the Atwood number and is defined as:
\begin{align}
    A = \frac{\rho_{hf}-\rho_{lf}}{\rho_{hf}+\rho_{lf}}
\end{align}
where $\rho_{hf} = \epsilon_f \rho_f + (1-\epsilon_f)\rho_p$ is the density of the heavy fluid since the heavy fluid is a mixture of fluid and particles and $\rho_{lf} = \rho_f$ is the density of the light fluid. The constant $\alpha$ varies slightly with the Atwood number. It is believed to be insensitive to initial conditions \cite{dimonte}. Therefore, we start all our simulations with the same initial conditions. According to Snider, it lies between 0.05 and 0.07 \cite{SniderD.M2001AITM}. A historical survey by Dimonte et al.\cite{dimonte} for $\alpha$ obtained from experiments and numerical simulations found that $\alpha$ can have values between 0.03 and 0.08 \cite{dimonte}. It is important to note that Eq. (\ref{eq::mixing_layer}) is only valid after the initial exponential growth period. We expect to obtain slightly larger values of $\alpha$ in our simulations as we are working in 3D where single modes grow faster than in 2D \cite{dimonte}.

\subsubsection{Simulation Setup}
Similar to the fluidized bed test case setup, we insert the particles using the DEM solver before starting the CFD-DEM simulation. Particles of the same diameter and density were non-uniformly and randomly inserted in the upper section (initial height of particles is 0.05 $m$) of a 3 dimensional rectangular bed. No initial perturbation is placed on the interface of the two fluids; however, small perturbations occur as a result of the random distribution of particles. We use the Dallavalle drag model \cite{dallavalle} as we want a drag force independent of the void fraction. We solve model A of the VANS equations. We apply slip boundary conditions on the side walls of the beds as to eliminate boundary effects. We apply no slip boundary conditions on the top and bottom walls to prevent fluid re-circulation. Three test cases were simulated with varying Atwood numbers. They are presented in Table \ref{tab:setup3}.

\begin{table}[H]
 \captionof{table}{Parameters for the setup of the Rayleigh-Taylor Instability.}
 \label{tab:setup3}
\centering
\resizebox{\textwidth}{!}{%
 \begin{tabular}{l l l l} 
 \hline
Particle radius ($m$) & 0.0001 & 0.0001 & 0.0001\\
Particle density ($kg/m^3$) & 3 & 5 & 10\\
Fluid density ($kg/m^3$) & 1 & 1 & 1\\
Initial particle volume fraction & 0.088 & 0.088 & 0.088\\
gravity ($m/s$) & 0,-9.8,0 & 0,-9.8,0 & 0,-9.8,0 \\
Number x,y,z mesh size & $18\times 150 \times 18$ & $18\times 150 \times 18$ & $18\times 150 \times 18$\\
Bed x,y,z size ($m$) & $0.01\times 0.15 \times 0.01$ &  $0.01\times 0.15 \times 0.01$ &  $0.01\times 0.15 \times 0.01$\\
Atwood Number & 0.0817 & 0.151 & 0.286\\
Number of Particles & 837925 & 837925 & 837925\\
CFD time step ($s$) & $5\times 10^{-4}$ & $5\times 10^{-4}$ & $5\times 10^{-4}$\\
CFD-DEM coupling frequency & 500 & 500 & 500\\
 \hline
 \end{tabular}}
\end{table}

Our choice of densities result in a close range of Atwood numbers, therefore, we expect that the difference in the instabilities and the mixing length will be relatively small.  

\subsubsection{Results and Discussion}
\begin{figure}[H]
        \centering
        \includegraphics[width=0.8\textwidth]{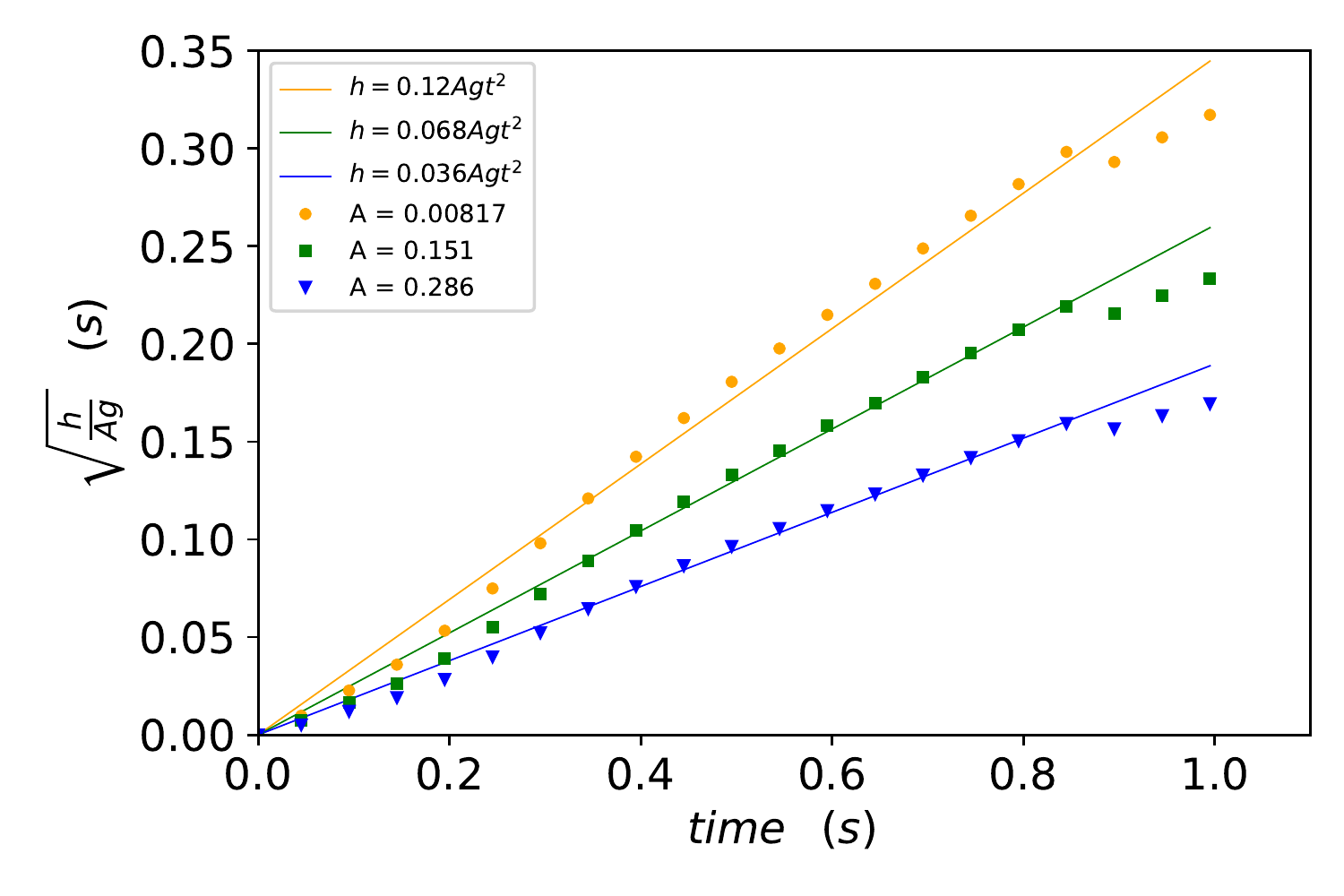}
        \caption{Growth rate of the mixing layer as a function of time.}
        \label{fig:mixing_layer}
\end{figure}

We measure the mixing width as the height of the bed starting from the first position where we have the diluted region of particles and ending with the last denser layer having less particles than the initial particle packing. This ensure we capture the complete mixing width. Fig. \ref{fig:mixing_layer} shows the plot of the growth rate of the mixing layer as a function of time. 

\begin{figure}[H]
        \centering
        \includegraphics[width=0.7\textwidth]{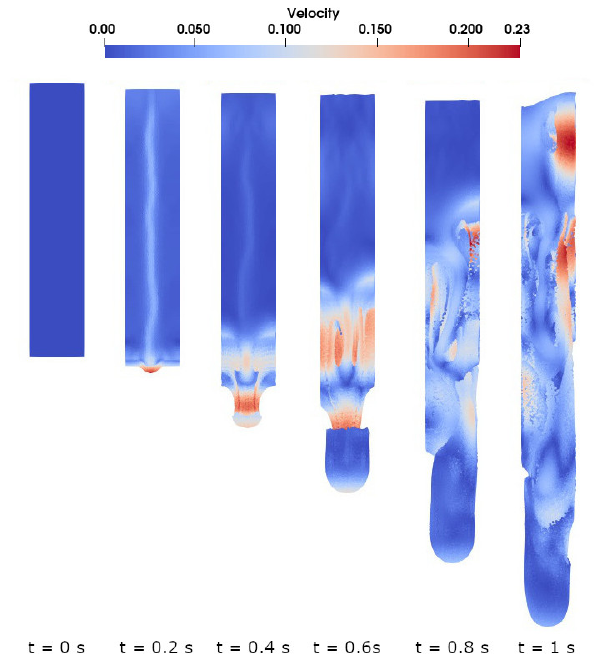}
        \caption{Side view of the 3D growing Rayleigh-Taylor instability plumes.}
        \label{fig:plumes}
\end{figure}

The measurement of the mixing layer agrees well with the correlation of the mixing layer. The initial $0.2$ $s$ do not agree with the correlation as they represent the initial exponential instability growth where correlation (\ref{eq::mixing_layer}) is not valid. Additionally, the sudden drop observed at around time $0.9$ $s$ is attributed to a big bubble of the light fluid reaching the top of the bed. As such, the packing at the top of the bed is disturbed, and not enough particles remain to feed the instability which eventually leads to the breaking of the mixing layer, thus rendering correlation (\ref{eq::mixing_layer}) invalid after $0.9$ $s$. We observe this behavior in Fig. \ref{fig:plumes} which shows snapshots depicting the growth of the Rayleigh-Taylor instability at different times in the simulation for $A = 0.286$. 

\begin{figure}[H]
        \centering
        \includegraphics[width=0.5\textwidth]{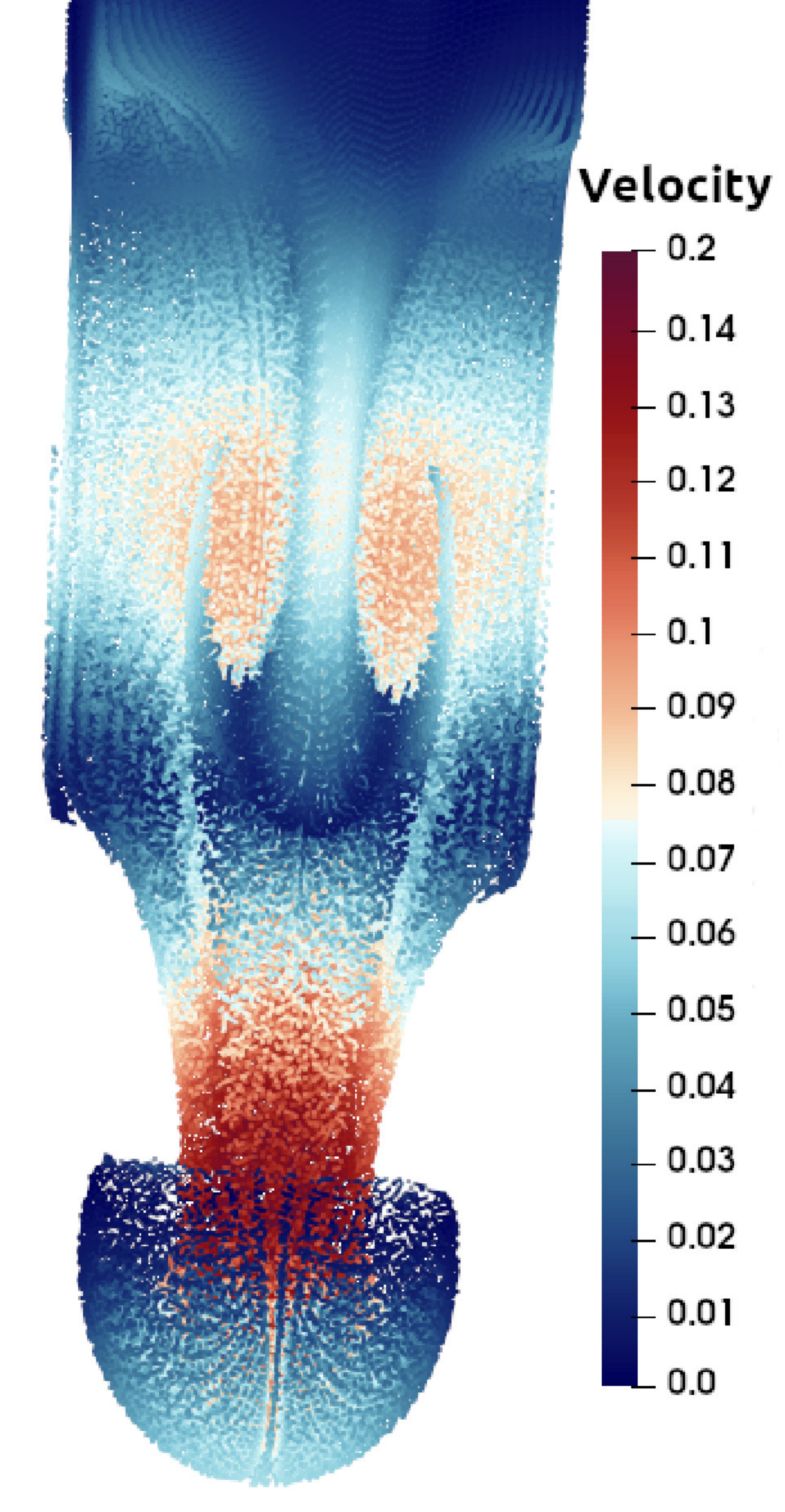}
        \caption{Angle view of the 3D growing Rayleigh-Taylor instability plumes for A=0.286 and t=0.45 s.}
        \label{fig:plumes_angle}
\end{figure}

The bubble at $t = 1$ $s$ is responsible for the sudden turn over in the curves observed at the end of the mixing layer growth of Fig. \ref{fig:mixing_layer}. From an angle, we show the 3D plumes for $A = 0.286$ at $t = 0.45$ $s$ in Fig. \ref{fig:plumes_angle} . 

The instability is formed as a major single central mushroom-shaped plume surrounded by minor plumes. The instability remains symmetrical until $t = 0.5$ $s$ after which it breaks down. The symmetrical plumes created demonstrate the mass conservation property of our solver. This is clear from the formation of the plumes where the light fluid moves upward to allow space for the heavy fluid moving downwards. Additionally, the circular shape of the mushroom inwards at its extremity is only achieved due to the highly accurate $Q_n$ interpolation of the particles' properties and position within a cell.

\subsection{Spouted Bed Test Case}
We simulate a rectangular spouted bed. We calculate the time averaged particles' velocities at different positions in the bed and compare the data obtained with experimental results obtained using particle image velocimetry \cite{YueYuanhe2020EIoS}. 

\subsubsection{Simulation Setup}
\begin{figure}[H]
        \centering
        \includegraphics[width=0.5\textwidth]{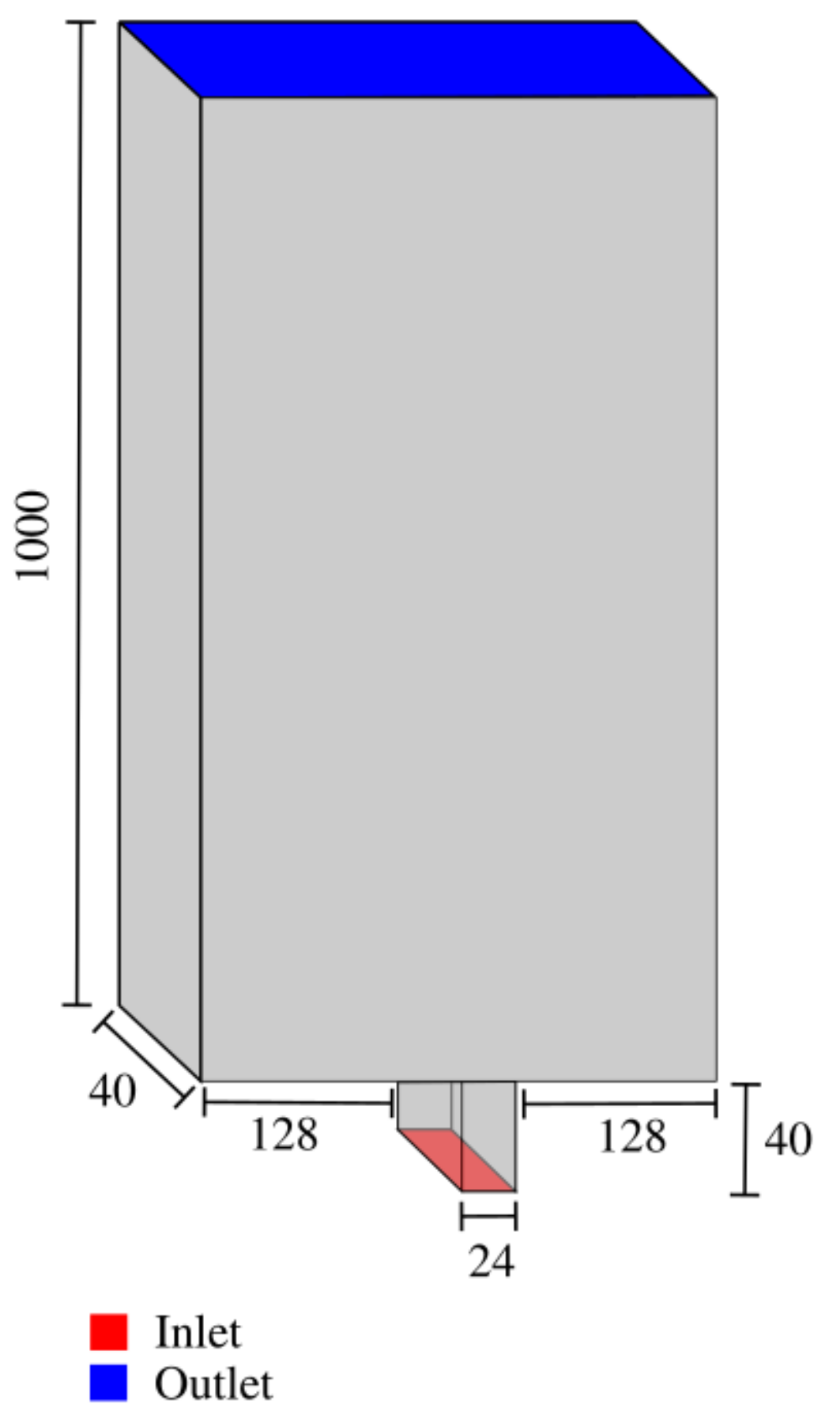}
        \caption{Scheme of the spouted bed geometry (unit: mm).}
        \label{fig:schematic_spouted_bed}
\end{figure}

The bed size, geometry, solid and fluid properties are extracted from the experiment of Yue et al. \cite{YueYuanhe2020EIoS}. Fig. \ref{fig:schematic_spouted_bed} shows the schematic geometry of the simulation. 

The authors did not specify the number of particles used but they specified the static bed height to be 200 $mm$. Based on the given geometry, and in order to obtain a similar static bed height, we use 175800 particles with the same properties as those used by Yue et al. \cite{YueYuanhe2020EIoS}. However, in order to obtain better distribution of the flow and stability of the solver, we add to the geometry a small channel at the bottom of the bed through which we introduce the gas. The geometry and mesh were generated using the open source software GMSH. For this case, we solve model A of the VANS equations. The simulation is performed twice using Q1-Q1 and Q2-Q1 elements, a smoothing length of $L^2 =  5d_p^2$ for the void fraction, and a grad-div stabilization parameter ($c^*$) equal to the element size. For the Q2-Q1 simulation, convergence was difficult for ($c^*$) the size of the element, so the value of ($c^*$) 10 times the element size was chosen as it made the Q2-Q1 simulations robust. We increment the spout inlet velocity gradually from $0$ $m/s$ at $ t = 0$ $s$ until it reached its maximum value of $20.8$ $m/s$ at $ t = 0.05$ $s$. This ensure a smooth transition of the bed from rest to the fluidized regime and prevents an initial shock from occurring due to the sudden introduction of a high velocity at $ t = 0$ $s$. Table \ref{tab:setup4} presents the physical and numerical parameters for this test case. 

\begin{table}[H]
 \captionof{table}{Physical and numerical parameters for the setup of the spouted bed test case.}
 \label{tab:setup4}
\centering
\resizebox{\textwidth}{!}{%
 \begin{tabular}{l l l l} 
 \hline
 \textit{Simulation control}\\
 End time ($s$) & 20 & Time discretization scheme & BDF2\\
Coupling frequency  &  $100$ & CFD time step ($s$) & $5 \times 10^{-4}$\\
 \\
 \textit{Geometry}\\
 Bed Height ($mm$) & 1000 &  Bed Width ($mm$) & 280 \\
 Bed Depth ($mm$) & 40 & Wall Thickness ($mm$) & 0 \\ 
 Channel Height ($mm$) & 40 &  Channel Width ($mm$) & 24 \\
 Channel Depth ($mm$) & 40 \\
 \\
 \textit{Mesh}\\
 Mesh Type & gmsh \\
 Bed Subdivisions in x-y-z & 36-100-4 & Channel Subdivisions in x-y-z & 2-5-4\\
 \\
 \textit{Particles}\\
 Number & $175800$ & Diameter(m) & $25\times10^{-4}$\\
 Density($kg/m^3$) & 2500 & Young Modulus ($N/m^2$) &  $1\times10^7$\\
 Particle-particle poisson ratio & 0.25 & Particle-wall poisson ratio & 0.25 \\
 Particle-particle restitution coefficient & 0.9 & Particle-wall restitution coefficient & 0.9 \\
 Particle-particle friction coefficient & 0.3 & Particle-wall friction coefficient & 0.3  \\
 Particle-particle rolling friction & 0.1 & Particle-wall rolling friction & 0.1 \\
 \\
 \textit{Gas phase}\\
 Viscosity ($Pa\cdot s$) & $1.81\times10^{-5}$ & Density ($kg/m^3$) & 1 \\
 Inlet Velocity ($m/s$) & 20.8 \\
 \\
 \textit{Linear Solver}\\
 Method & GMRES & Max iterations & 5000\\
 Minimum residual & $10^{-8}$ & Relative residual & $10^{-2}$\\
 ILU preconditioner fill & 1 & ILU preconditioner absolute tolerance & $10^{-10}$\\
 ILU preconditioner relative tolerance & 1 & max krylov vectors & 200\\
 \\
 \textit{Non-linear solver}\\
 Tolerance & $1\times10^{-9}$ & Max iterations & 25\\
  \hline
 \end{tabular}}
\end{table}

We perform the simulations for four particle-particle and particle-wall friction coefficients (0.3, 0.2, 0.1, and 0.05) and determine that a value of 0.3 for both allowed for the most realistic behavior of the spouted bed that is comparable to the behavior obtained from the experiments. The coefficient of rolling friction had minor or no effects on the behavior of the bed and was thus kept fixed at 0.1.  We apply slip boundary condition for all walls of the bed and the channel except the channel's base (inlet) and the bed's top wall (outlet). At the inlet, we apply a Dirichlet boundary condition with the value of the inlet velocity in the y-direction. For the outlet, we apply a zero traction boundary condition when the fluid is leaving the domain and we penalize the flow when the fluid is inbound. This prevents flow re-entry when there is turbulent structures or vortices leaving the domain. The outlet boundary condition imposed is:

\begin{align}
    \int_{\Gamma_o}(\nu \nabla \bm{u} \cdot \bm{n} - \emph{p I} \cdot \bm{n} - \beta(\bm{u} \cdot \bm{n})\_\bm{u}) \cdot \bm{v} \, d\Gamma = 0
\end{align}

where $\beta$ is a constant, $\emph{p}$ is the pressure, $\emph{I}$ is the identity matrix, $(\bm{u} \cdot \bm{n})\_$ is $min(0,\bm{u} \cdot \bm{n})$ $\bm{v}$ is the velocity test function and $\Gamma_o$ is the outlet boundary. For further details, we refer the reader to the work of Arndt et al. \cite{ArndtDaniel2016FEft}. In order to prevent the particles from falling into the channel, we define a floating wall which is an imaginary wall only felt by the particles at the top of the channel at the intersection between the channel and the bed base.

\subsubsection{Results and Discussion}

\begin{figure}[H]
     \centering
     \begin{subfigure}[b]{0.75\textwidth}
         \centering
         \includegraphics[width=\textwidth]{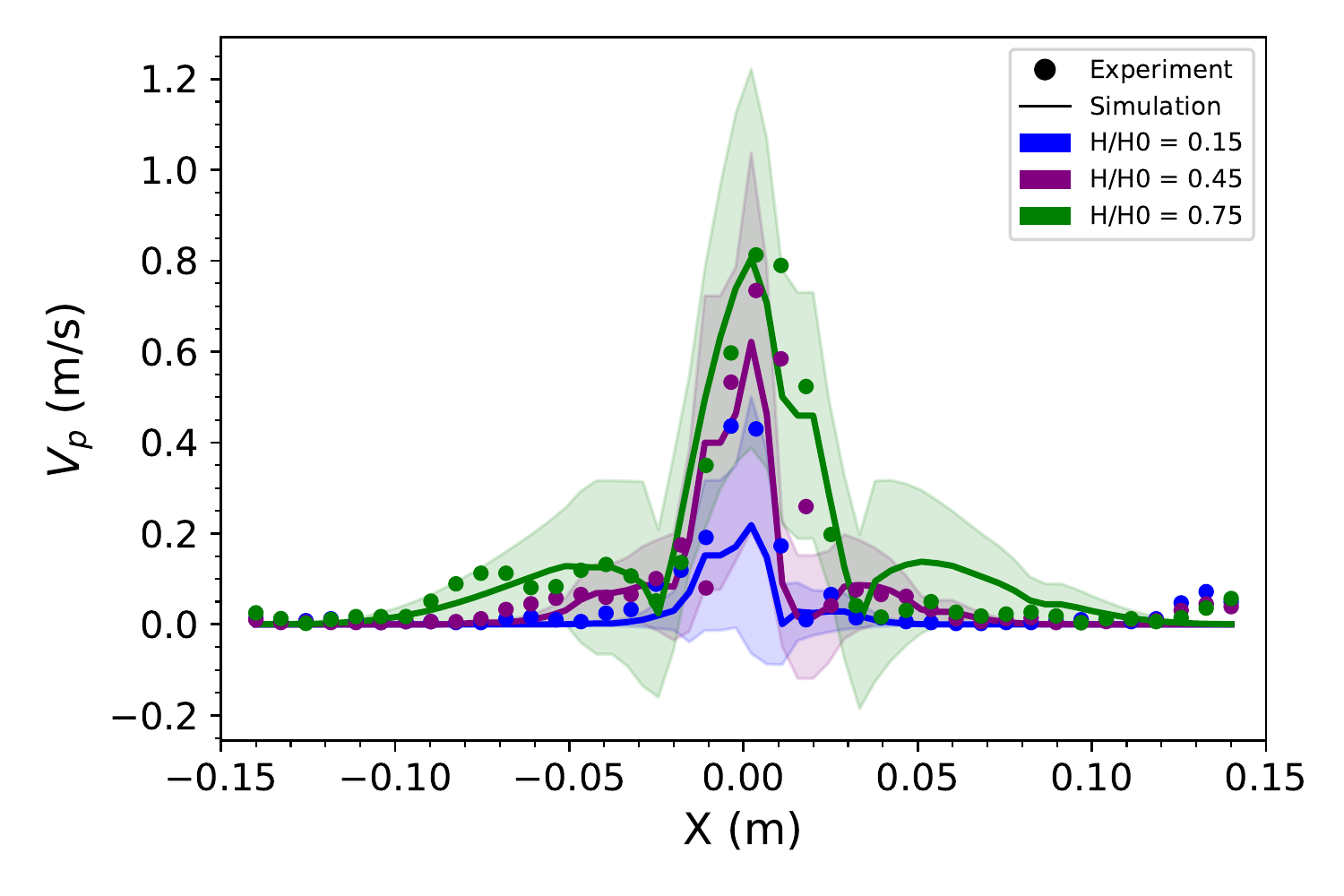}
         \label{fig:Q1_particle_velocity_1}
     \end{subfigure}
     \\
     \begin{subfigure}[b]{0.75\textwidth}
         \centering
         \includegraphics[width=\textwidth]{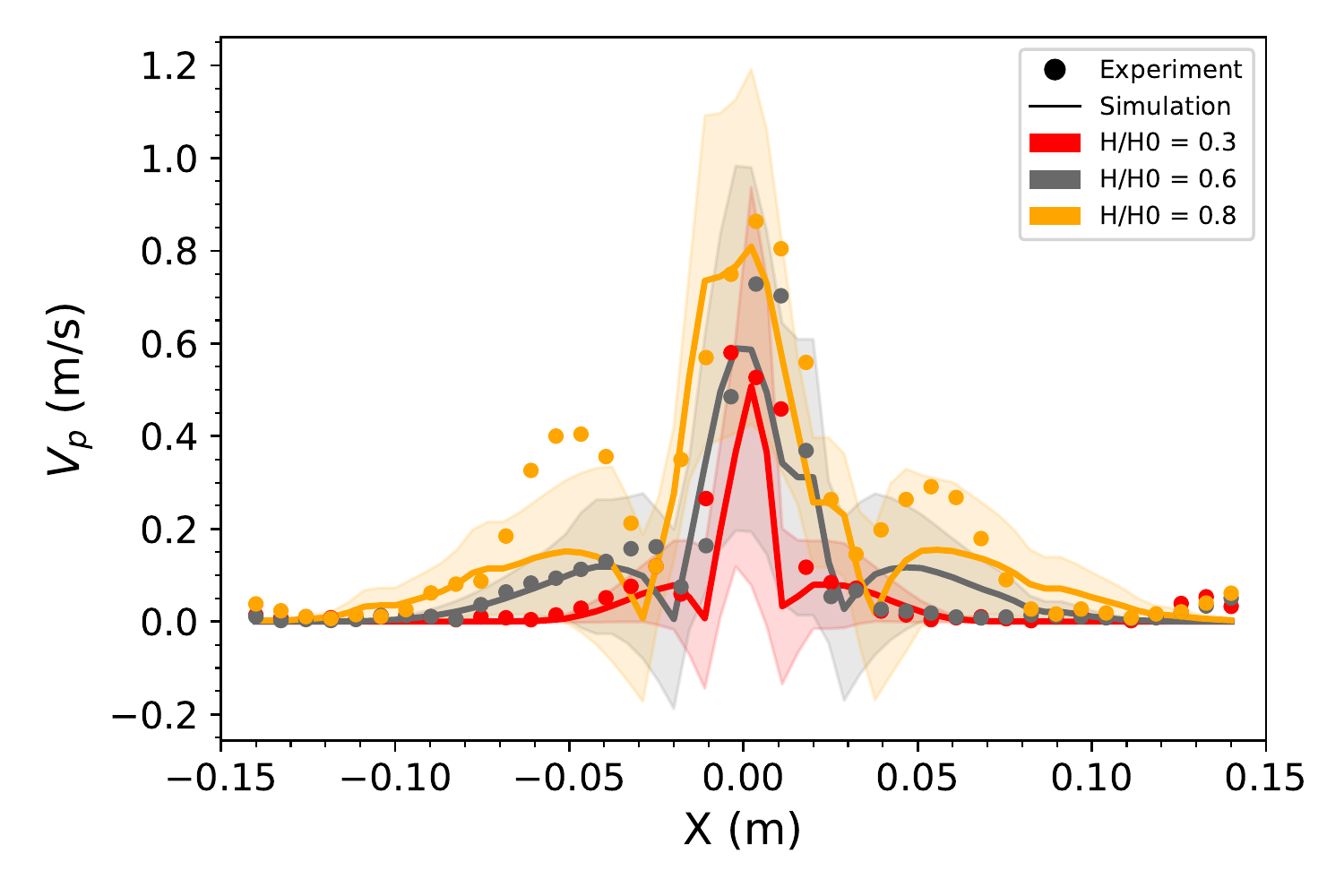}
         \label{fig:Q1_particle_velocity_2}
     \end{subfigure}
        \caption{Distribution of particle velocity magnitude in vectors along the X direction at different height levels for Q1-Q1 simulation.}
        \label{fig:Q1_particle_velocity}
\end{figure}

\begin{figure}[H]
     \centering
     \begin{subfigure}[b]{0.75\textwidth}
         \centering
         \includegraphics[width=\textwidth]{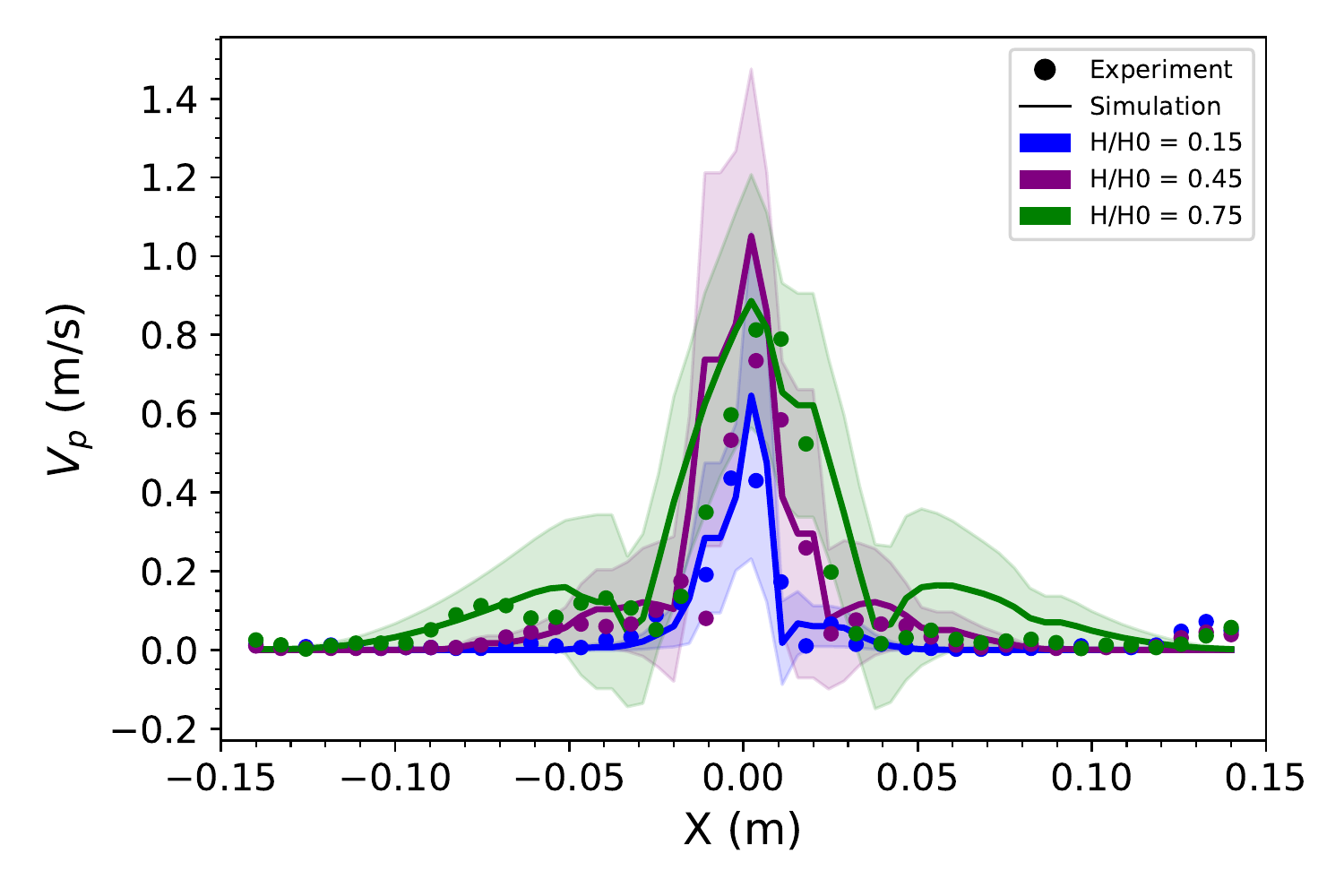}
         \label{fig:Q2_particle_velocity_1}
     \end{subfigure}
     \\
     \begin{subfigure}[b]{0.75\textwidth}
         \centering
         \includegraphics[width=\textwidth]{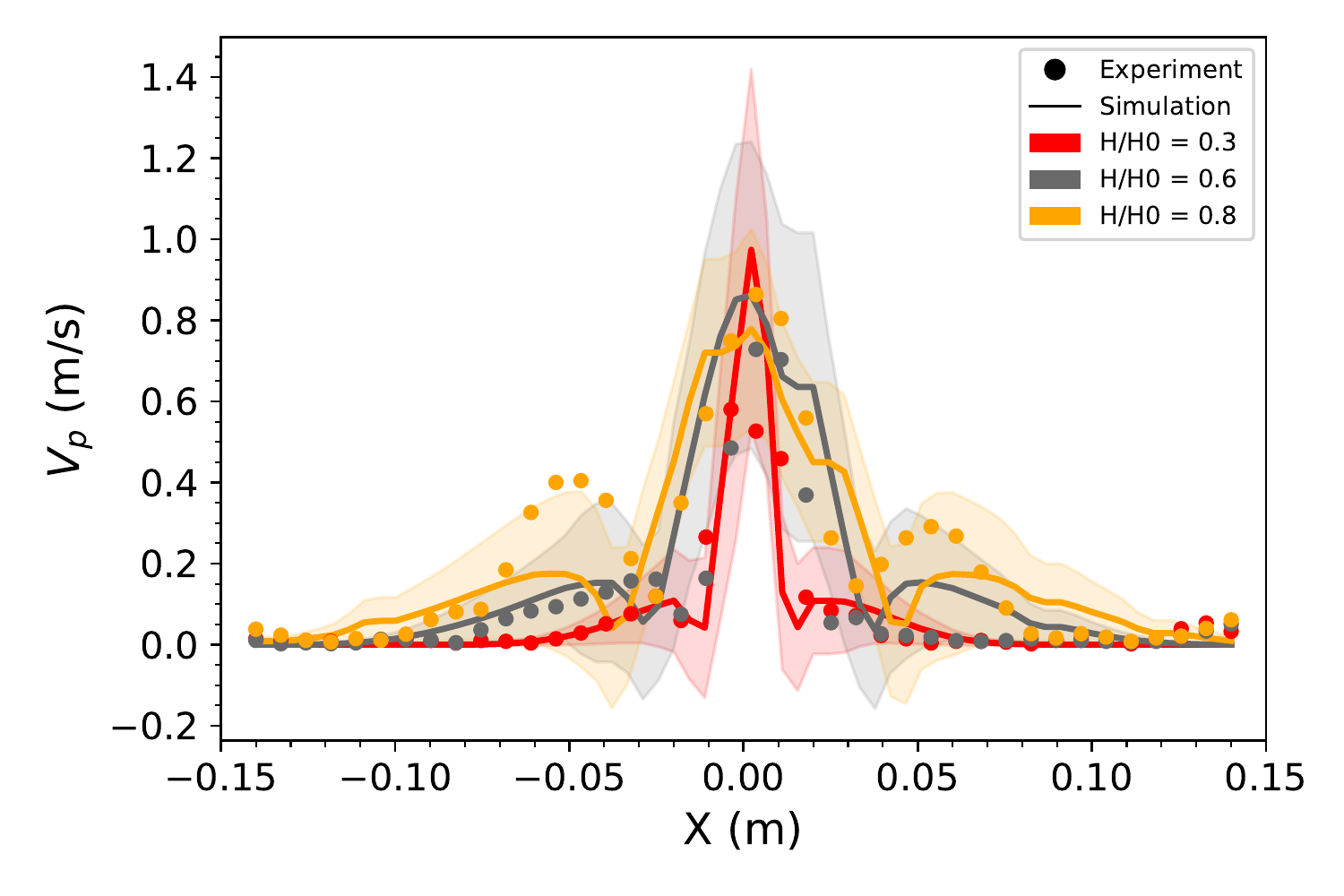}
         \label{fig:Q2_particle_velocity_2}
     \end{subfigure}
        \caption{Distribution of particle velocity magnitude in vectors along the X direction at different height levels for Q2-Q1 simulation.}
        \label{fig:Q2_particle_velocity}
\end{figure}

For both Q1-Q1 and Q2-Q1 simulations, we measure the time averaged particle velocity in the direction of the flow at different heights (y-axis) and widths (x-axis) of the bed and we compare the values with the experimental results of Yue et al . \cite{YueYuanhe2020EIoS}. The heights in Figs . \ref{fig:Q1_particle_velocity} and \ref{fig:Q2_particle_velocity} are normalized by the static bed height.

Fig . \ref{fig:Q1_particle_velocity} shows the average particles' velocity magnitude at different heights in the spouted bed along the bed's width for the Q1-Q1 simulation. The spout is centralized at $x = 0$ $m$ and the velocities obtained from the simulation agrees well with the experimental velocities. It is important to note that the experimental velocities were measured using Particle Image Velocimetry (PIV) which already involves errors in measurement. Around the central spout, we observe re-circulation zones at $ x \in [0.05,0.1] \: m \And  x \in [-0.1,-0.05] \: m$. The calculated re-circulation velocity magnitude is smaller than that measured in the experiments. This can be due to errors in the PIV measurements as the zones are not symmetric in the experiments contrary to what it should be. Additionally, we calculate the particles' average velocity magnitude by averaging over a finite element which is different than the averaging procedure performed in the experiments. However, the general trend of the results follows well that of the experiments. We average over the last 15 $s$ out of the 20 $s$ of the simulation to ensure that the averaging is performed only when the pseudo steady state has been reached. We show the standard deviation of our curves as a zone of the same color. This standard deviation represents the fluctuation of particles' velocity at this location with respect to time. All experimental velocities lie in the respective standard deviation zones except for the re-circulation velocities at $H/H_0 = 0.8$. This demonstrates that despite all possible measurement and averaging errors, the simulation gives good results comparable to the experimental results of the bed.

To investigate and better understand the effect of high order methods, we simulated the same case using Q2-Q1 elements. The results are shown in Fig. \ref{fig:Q2_particle_velocity} 

The differences between Q1-Q1 and Q2-Q1 simulations' results are mainly localized in the spout region the closer our reference height is to inlet of the bed. The four smallest heights from 0.15 till 0.6 have a much higher average particle velocity compared to the Q1-Q1 simulation while the two largest heights of 0.75 and 0.8 have a relatively similar velocity profile in both simulations. This can be attributed to the presence of turbulent structures at the inlet due to the sudden shock experienced by the fluid traveling from a narrow to a wide region causing an unstable high velocity gradient. These turbulent structures which in general possess a greater kinetic energy are better captured using higher order elements leading to an increase in the particles' average velocities in this region. In general, the oscillation of the curves in time are more prominent in the Q1-Q1 than in the Q2-Q1 results. This explains the overall thinner standard deviation zones in Fig. \ref{fig:Q2_particle_velocity} compared to Fig. \ref{fig:Q1_particle_velocity}. However, around $x = 0$ $m$ and for the heights closer to the inlet, the standard deviation is high. This is caused by the high velocity and turbulent structures in this region leading to strong oscillations with time. Finally, the experimental particles' velocities in the re-circulation zones for height 0.8 lies within the standard deviation zone calculated from the simulation. Thus, this zone is better captured in the Q2-Q1 simulation.

\subsubsection{Effect of load balancing on computational time}
\begin{figure}[H]
        \centering
        \includegraphics[width=0.9\textwidth]{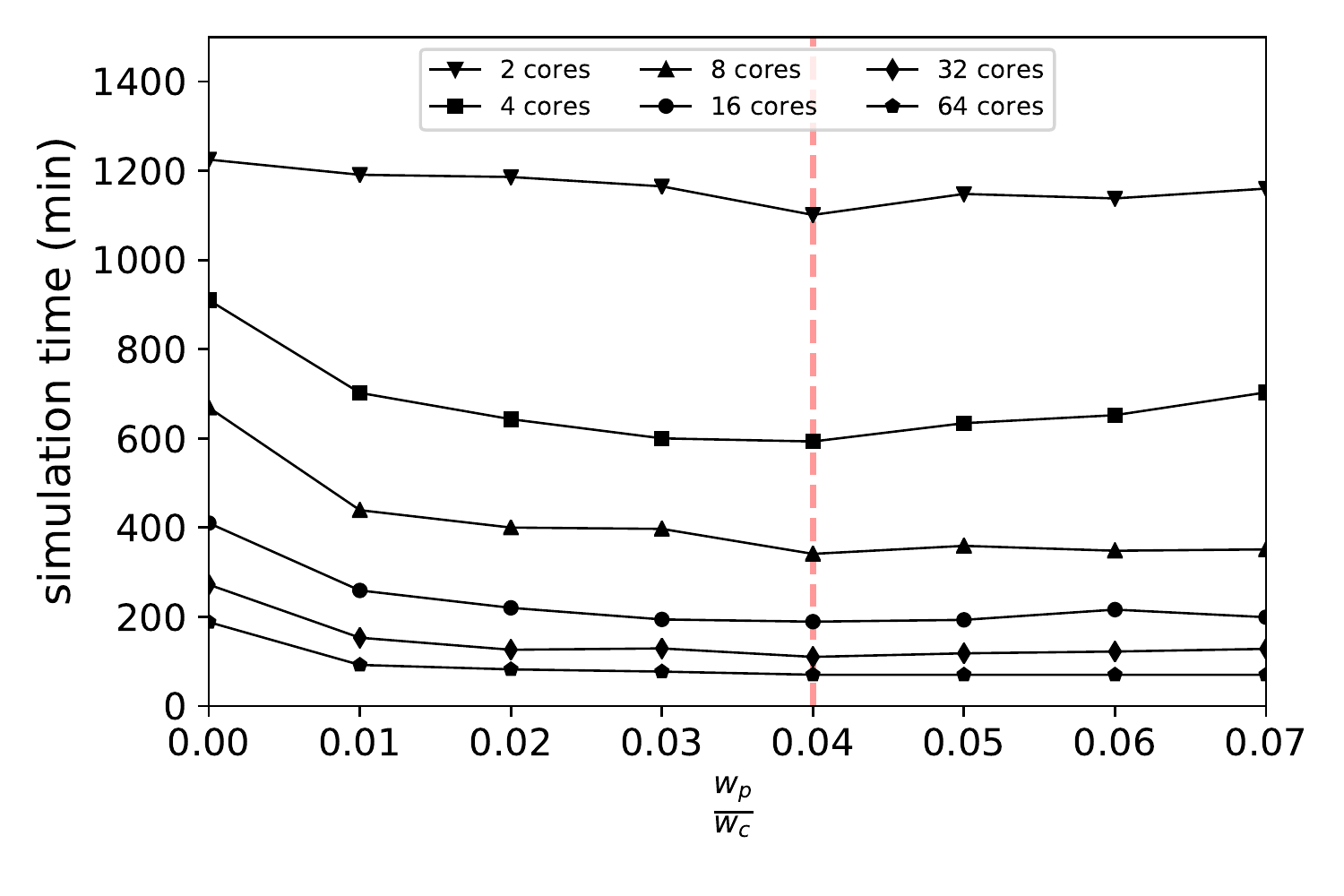}
        \caption{Overall simulation time as a function of the the particle to cell weight ratio for different number of processors.}
        \label{fig:simulation_time}
\end{figure}

Our CFD-DEM solver supports load balancing; however, there are many parameters that should be accounted for to ensure enhanced computational efficiency such as the ratio of solid particle to fluid cell weight and the frequency of load balancing. Among these parameters, we focus on studying the ratio of solid particle to fluid cell weight. For the purpose of this study and to reduce the computational cost of the simulation, we simulated the spouted bed test case with a coarser mesh of 18-70-4 for the bed and 2-3-4 for the channel in x-y-z and with half the number of particles (87,900 particles) using Q1-Q1 elements for different particle to fluid cell weight ratios at six different numbers of processors. The simulation was run for $1$ $s$ of real time and a time step of $0.001$ $s$ using the dynamic load balancing approach which automatically detects the load balancing steps from the distribution of particles and cells among the processors \cite{GolshanShahab2021L:Ao}. All other parameters were kept the same as given in Table \ref{tab:setup4}. The result is shown in Fig. \ref{fig:simulation_time}.

For all simulations, a particle to cell weight ratio of 0.04 is the optimum ratio. However, the difference in speed between two neighboring ratios of 0.04 is negligible as it only adds few minutes to the simulation. As such, knowing the neighborhood of the optimal weight ratio to use is sufficient as it can affect drastically the speed especially for simulations with a lower number of processors. Moreover, looking at the trend of simulation time as a function of weight ratios, we observe that the bigger the number of processors used, the less effect the change in weight ratios has on the overall simulation time. 

\begin{figure}[H]
        \centering
        \includegraphics[width=0.9\textwidth]{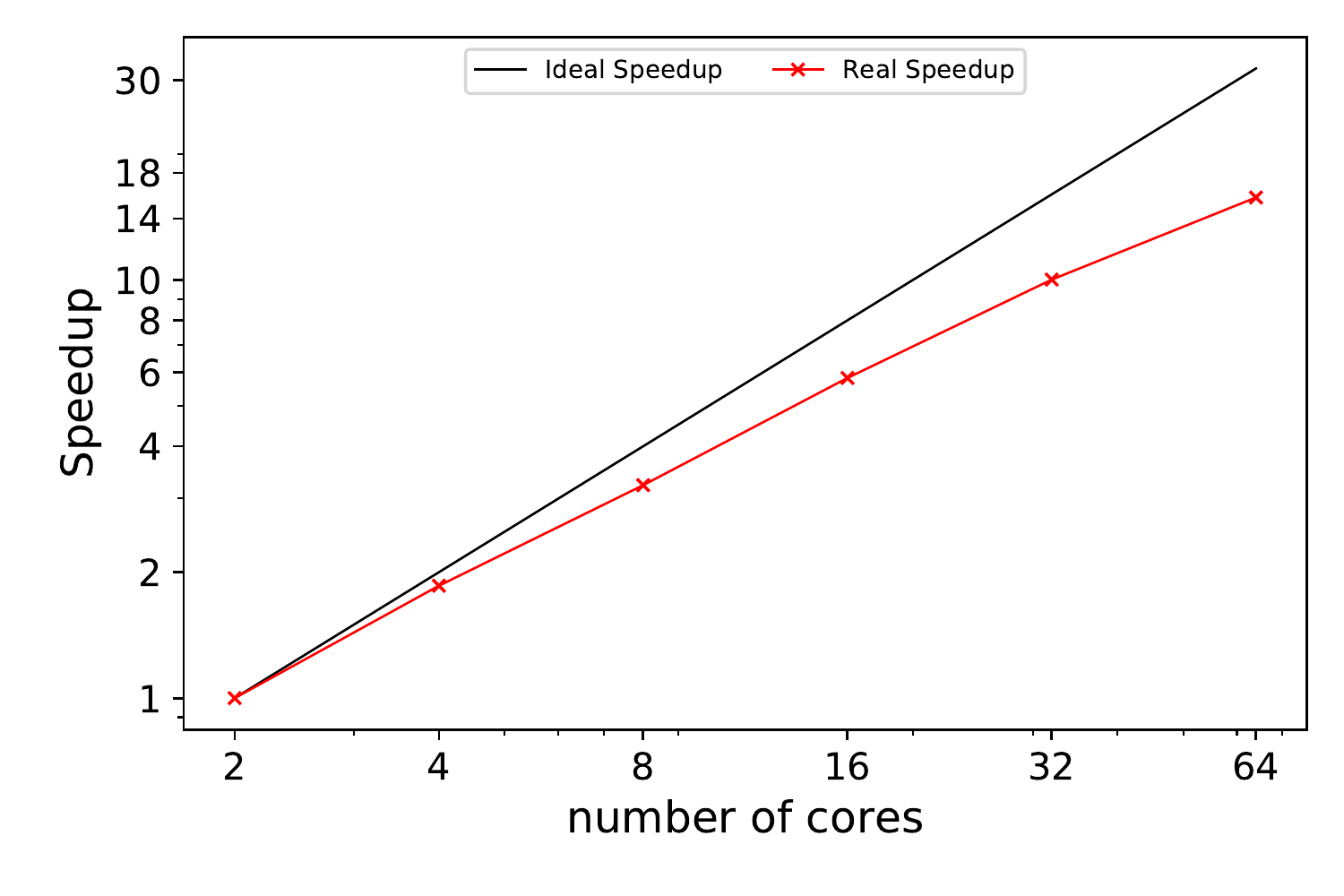}
        \caption{Simulation speedup as a function of processors in logarithmic scale}
        \label{fig:simulation_speedup}
\end{figure}

We plot the speedup of the simulation as a function of the number of processors in a logarithmic scale as shown in Fig. \ref{fig:simulation_speedup}.

Overall, the CFD-DEM solver scales well. For a lower number of processors, our solver scales better than for an increased number of processors. This is because at lower number of processors, the matrix assembly of the CFD solver is the most consuming part and takes around 70 \% of a time step. The higher the number of processors, the faster the matrix assembly gets as the matrices on each processor become smaller. However, increasing the number of processors beyond a certain value, in this case 16, leads to a decrease in the scaling. This is due to the DEM solver that becomes slower. Our DEM solver uses the notion of ghost particles which are particles seen by the current processor but owned by another neighboring processor. As the number of processors increase, the number of ghost particles increase as well as a single particle is owned by one processor but can be seen as a ghost particle by more and more processors. This leads to an increase of the computational cost as the DEM solver has to perform more calculations related to the ghost particles. For 2 processors, an average of 43950 particles exists in a single processor. This value decreases by a factor of 2 everytime we double the number of processors. At 16 processors, this value is around 5493 particles per processor. Additional decrease in this value leads to a decrease in the scalability of the DEM solver for which quasi-ideal scaling could be obtained at 20,000 particles per core \cite{GolshanShahab2021L:Ao}.

\section{Conclusion}
This work presents a verified and validated stabilized finite element approach for the modeling of multi-phase flows using unresolved CFD-DEM. Our monolithic coupled solver is robust, parallel and supports both models A and B of the VANS equations. It supports high order finite elements which allows for more accurate results using bigger mesh sizes thus respecting the requirement for large enough mesh sizes without compromising accuracy. Also, it is among the first CFD-DEM software to support load balancing. This helps improve the computational efficiency of our solver as available resources can be exploited more effectively by balancing work loads among processors. Moreover, the DEM and CFD portions of the code run on the same mesh which results in cheaper communication and coupling between them.  Furthermore, our formulation is locally and globally conservative.

Our solver can simulate solid-fluid flows for different applications. We validate the code with different test cases. In the fluidized bed test case, we show that even though model A and B are mathematically equivalent, they result in different instantaneous physical behavior of fluidized beds. Model A is more stable and results in less stiffness of the system being solved. From the Rayleigh Taylor instability, we prove mass conservation of our solver as well as accurate interpolation of particle properties. The spouted bed test case allows us to compare the particles' velocity at different bed heights with experimental data. We can thus conclude that our results are realistic and our formulation is validated.


\section{Acknowledgements}
The authors would like to thank the deal.II community for their support as well as Professor Yansong Shen for his clarification concerning the spouted bed experimental results. Bruno Blais would like to acknowledge the financial support from the Natural Sciences and Engineering Research Council of Canada (NSERC) through the RGPIN-2020-04510 Discovery Grant. The authors would also like to acknowledge technical support and computing time provided by Compute Canada and Calcul Québec.

\newpage

\bibliography{library}

\providecommand{\latin}[1]{#1}
\makeatletter
\providecommand{\doi}
  {\begingroup\let\do\@makeother\dospecials
  \catcode`\{=1 \catcode`\}=2 \doi@aux}
\providecommand{\doi@aux}[1]{\endgroup\texttt{#1}}
\makeatother
\providecommand*\mcitethebibliography{\thebibliography}
\csname @ifundefined\endcsname{endmcitethebibliography}
  {\let\endmcitethebibliography\endthebibliography}{}
\begin{mcitethebibliography}{47}
\providecommand*\natexlab[1]{#1}
\providecommand*\mciteSetBstSublistMode[1]{}
\providecommand*\mciteSetBstMaxWidthForm[2]{}
\providecommand*\mciteBstWouldAddEndPuncttrue
  {\def\EndOfBibitem{\unskip.}}
\providecommand*\mciteBstWouldAddEndPunctfalse
  {\let\EndOfBibitem\relax}
\providecommand*\mciteSetBstMidEndSepPunct[3]{}
\providecommand*\mciteSetBstSublistLabelBeginEnd[3]{}
\providecommand*\EndOfBibitem{}
\mciteSetBstSublistMode{f}
\mciteSetBstMaxWidthForm{subitem}{(\alph{mcitesubitemcount})}
\mciteSetBstSublistLabelBeginEnd
  {\mcitemaxwidthsubitemform\space}
  {\relax}
  {\relax}

\bibitem[Yeoh and Tu(2019)Yeoh, and Tu]{ch4}
Yeoh,~G.~H.; Tu,~J. In \emph{Computational Techniques for Multiphase Flows
  (Second Edition)}, second edition ed.; Yeoh,~G.~H., Tu,~J., Eds.;
  Butterworth-Heinemann, 2019; pp 227--319\relax
\mciteBstWouldAddEndPuncttrue
\mciteSetBstMidEndSepPunct{\mcitedefaultmidpunct}
{\mcitedefaultendpunct}{\mcitedefaultseppunct}\relax
\EndOfBibitem
\bibitem[Norouzi.(2016)]{norouzi2016}
Norouzi.,~H.~R. \emph{Coupled CFD-DEM Modeling : Formulation, Implementation
  and Application to Multiphase Flows.}, 1st ed.; Wiley: Newark, 2016\relax
\mciteBstWouldAddEndPuncttrue
\mciteSetBstMidEndSepPunct{\mcitedefaultmidpunct}
{\mcitedefaultendpunct}{\mcitedefaultseppunct}\relax
\EndOfBibitem
\bibitem[Yeoh and Tu(2019)Yeoh, and Tu]{ch1}
Yeoh,~G.~H.; Tu,~J. In \emph{Computational Techniques for Multiphase Flows
  (Second Edition)}, second edition ed.; Yeoh,~G.~H., Tu,~J., Eds.;
  Butterworth-Heinemann, 2019; pp 1 -- 18\relax
\mciteBstWouldAddEndPuncttrue
\mciteSetBstMidEndSepPunct{\mcitedefaultmidpunct}
{\mcitedefaultendpunct}{\mcitedefaultseppunct}\relax
\EndOfBibitem
\bibitem[van~der Hoef \latin{et~al.}(2006)van~der Hoef, Ye, van Sint~Annaland,
  Andrews, Sundaresan, and Kuipers]{vanderHoefM.A2006MMoG}
van~der Hoef,~M.; Ye,~M.; van Sint~Annaland,~M.; Andrews,~A.; Sundaresan,~S.;
  Kuipers,~J. Multiscale Modeling of Gas-Fluidized Beds. \emph{Advances in
  Chemical Engineering} \textbf{2006}, \emph{31}, 65--149\relax
\mciteBstWouldAddEndPuncttrue
\mciteSetBstMidEndSepPunct{\mcitedefaultmidpunct}
{\mcitedefaultendpunct}{\mcitedefaultseppunct}\relax
\EndOfBibitem
\bibitem[Sahoo and Sahoo(2013)Sahoo, and Sahoo]{SahooPranati2013FaSo}
Sahoo,~P.; Sahoo,~A. Fluidization and Spouting of Fine Particles: A Comparison.
  \emph{Advances in materials science and engineering} \textbf{2013},
  \emph{2013}, 1--7\relax
\mciteBstWouldAddEndPuncttrue
\mciteSetBstMidEndSepPunct{\mcitedefaultmidpunct}
{\mcitedefaultendpunct}{\mcitedefaultseppunct}\relax
\EndOfBibitem
\bibitem[Cocco \latin{et~al.}(2014)Cocco, Karri, and Knowlton]{CoccoRay2014ItF}
Cocco,~R.; Karri,~S. B.~R.; Knowlton,~T. Introduction to Fluidization.
  \emph{Chemical engineering progress} \textbf{2014}, \emph{110}, 21--29\relax
\mciteBstWouldAddEndPuncttrue
\mciteSetBstMidEndSepPunct{\mcitedefaultmidpunct}
{\mcitedefaultendpunct}{\mcitedefaultseppunct}\relax
\EndOfBibitem
\bibitem[ede()]{edem}
Discrete Element Modeling - DEM Software: Altair Edem.
  \url{https://www.altair.com/edem/}\relax
\mciteBstWouldAddEndPuncttrue
\mciteSetBstMidEndSepPunct{\mcitedefaultmidpunct}
{\mcitedefaultendpunct}{\mcitedefaultseppunct}\relax
\EndOfBibitem
\bibitem[roc(2022)]{rocky}
Rocky DEM. 2022; \url{https://rocky.esss.co/software/}\relax
\mciteBstWouldAddEndPuncttrue
\mciteSetBstMidEndSepPunct{\mcitedefaultmidpunct}
{\mcitedefaultendpunct}{\mcitedefaultseppunct}\relax
\EndOfBibitem
\bibitem[Kloss \latin{et~al.}(2012)Kloss, Goniva, Hager, Amberger, and
  Pirker]{liggghts}
Kloss,~C.; Goniva,~C.; Hager,~A.; Amberger,~S.; Pirker,~S. Models, algorithms
  and validation for opensource DEM and CFD–DEM. \emph{Progress in
  Computational Fluid Dynamics, an International Journal} \textbf{2012},
  \emph{12}, 140--152\relax
\mciteBstWouldAddEndPuncttrue
\mciteSetBstMidEndSepPunct{\mcitedefaultmidpunct}
{\mcitedefaultendpunct}{\mcitedefaultseppunct}\relax
\EndOfBibitem
\bibitem[Smilauer \latin{et~al.}(2021)Smilauer, Angelidakis, Catalano, Caulk,
  Chareyre, Chèvremont, Dorofeenko, Duriez, Dyck, Elias, Er, Eulitz, Gladky,
  Guo, Jakob, Kneib, Kozicki, Marzougui, Maurin, Modenese, Pekmezi, Scholtès,
  Sibille, Stransky, Sweijen, Thoeni, and Yuan]{yade}
Smilauer,~V. \latin{et~al.}  \emph{Yade documentation}; The Yade Project,
  2021\relax
\mciteBstWouldAddEndPuncttrue
\mciteSetBstMidEndSepPunct{\mcitedefaultmidpunct}
{\mcitedefaultendpunct}{\mcitedefaultseppunct}\relax
\EndOfBibitem
\bibitem[com()]{comsol}
Comsol: a multiphysics software for optimizing designs.
  \url{https://www.comsol.com/}\relax
\mciteBstWouldAddEndPuncttrue
\mciteSetBstMidEndSepPunct{\mcitedefaultmidpunct}
{\mcitedefaultendpunct}{\mcitedefaultseppunct}\relax
\EndOfBibitem
\bibitem[ans()]{ansys}
Ansys FLUENT. \url{https://www.ansys.com/products/fluids/ansys-fluent}\relax
\mciteBstWouldAddEndPuncttrue
\mciteSetBstMidEndSepPunct{\mcitedefaultmidpunct}
{\mcitedefaultendpunct}{\mcitedefaultseppunct}\relax
\EndOfBibitem
\bibitem[sta()]{starccm}
Multiphysics Computational Fluid Dynamics (CFD) Simulation Software: Siemens
  Software.
  \url{https://www.plm.automation.siemens.com/global/en/products/simcenter/STAR-CCM.html}\relax
\mciteBstWouldAddEndPuncttrue
\mciteSetBstMidEndSepPunct{\mcitedefaultmidpunct}
{\mcitedefaultendpunct}{\mcitedefaultseppunct}\relax
\EndOfBibitem
\bibitem[Syamlal \latin{et~al.}(1993)Syamlal, Rogers, and OBrien]{mfix}
Syamlal,~M.; Rogers,~W.; OBrien,~T.~J. \emph{MFIX documentation theory guide};
  1993\relax
\mciteBstWouldAddEndPuncttrue
\mciteSetBstMidEndSepPunct{\mcitedefaultmidpunct}
{\mcitedefaultendpunct}{\mcitedefaultseppunct}\relax
\EndOfBibitem
\bibitem[ope()]{openfoam}
OpenFOAM. \url{https://www.openfoam.com/}\relax
\mciteBstWouldAddEndPuncttrue
\mciteSetBstMidEndSepPunct{\mcitedefaultmidpunct}
{\mcitedefaultendpunct}{\mcitedefaultseppunct}\relax
\EndOfBibitem
\bibitem[pfc()]{pfc}
PFC. \url{http://www.itascacg.com/software/pfc}\relax
\mciteBstWouldAddEndPuncttrue
\mciteSetBstMidEndSepPunct{\mcitedefaultmidpunct}
{\mcitedefaultendpunct}{\mcitedefaultseppunct}\relax
\EndOfBibitem
\bibitem[Golshan \latin{et~al.}(2022)Golshan, Munch, Gassm{\"o}ller,
  Kronbichler, and Blais]{GolshanShahab2021L:Ao}
Golshan,~S.; Munch,~P.; Gassm{\"o}ller,~R.; Kronbichler,~M.; Blais,~B.
  Lethe-DEM: An open-source parallel discrete element solver with load
  balancing. \emph{Computational Particle Mechanics} \textbf{2022}, 1--20\relax
\mciteBstWouldAddEndPuncttrue
\mciteSetBstMidEndSepPunct{\mcitedefaultmidpunct}
{\mcitedefaultendpunct}{\mcitedefaultseppunct}\relax
\EndOfBibitem
\bibitem[Blais \latin{et~al.}(2020)Blais, Barbeau, Bibeau, Gauvin, Geitani,
  Golshan, Kamble, Mirakhori, and Chaouki]{BlaisBruno2020LAop}
Blais,~B.; Barbeau,~L.; Bibeau,~V.; Gauvin,~S.; Geitani,~T.~E.; Golshan,~S.;
  Kamble,~R.; Mirakhori,~G.; Chaouki,~J. Lethe: An open-source parallel
  high-order adaptative CFD solver for incompressible flows. \emph{SoftwareX}
  \textbf{2020}, \emph{12}, 100579--\relax
\mciteBstWouldAddEndPuncttrue
\mciteSetBstMidEndSepPunct{\mcitedefaultmidpunct}
{\mcitedefaultendpunct}{\mcitedefaultseppunct}\relax
\EndOfBibitem
\bibitem[Arndt \latin{et~al.}(2021)Arndt, Bangerth, Blais, Fehling,
  Gassm{\"o}ller, Heister, Heltai, K{\"o}cher, Kronbichler, Maier, Munch,
  Pelteret, Proell, Simon, Turcksin, Wells, and Zhang]{dealII93}
Arndt,~D. \latin{et~al.}  The \texttt{deal.II} Library, Version 9.3.
  \emph{Journal of Numerical Mathematics} \textbf{2021}, \emph{29},
  171--186\relax
\mciteBstWouldAddEndPuncttrue
\mciteSetBstMidEndSepPunct{\mcitedefaultmidpunct}
{\mcitedefaultendpunct}{\mcitedefaultseppunct}\relax
\EndOfBibitem
\bibitem[Geitani \latin{et~al.}(2022)Geitani, Golshan, and
  Blais]{geitani2022high}
Geitani,~T.~E.; Golshan,~S.; Blais,~B. A High Order Stabilized Solver for the
  Volume Averaged Navier-Stokes Equations. \emph{arXiv preprint
  arXiv:2206.02842} \textbf{2022}, \relax
\mciteBstWouldAddEndPunctfalse
\mciteSetBstMidEndSepPunct{\mcitedefaultmidpunct}
{}{\mcitedefaultseppunct}\relax
\EndOfBibitem
\bibitem[GIDASPOW(1994)]{GIDASPOW19941}
GIDASPOW,~D. In \emph{Multiphase Flow and Fluidization}; GIDASPOW,~D., Ed.;
  Academic Press: San Diego, 1994; pp 1--29\relax
\mciteBstWouldAddEndPuncttrue
\mciteSetBstMidEndSepPunct{\mcitedefaultmidpunct}
{\mcitedefaultendpunct}{\mcitedefaultseppunct}\relax
\EndOfBibitem
\bibitem[Blais and Bertrand(2015)Blais, and Bertrand]{BlaisBruno2015Otuo}
Blais,~B.; Bertrand,~F. On the use of the method of manufactured solutions for
  the verification of CFD codes for the volume-averaged Navier–Stokes
  equations. \emph{Computers \& fluids} \textbf{2015}, \emph{114},
  121--129\relax
\mciteBstWouldAddEndPuncttrue
\mciteSetBstMidEndSepPunct{\mcitedefaultmidpunct}
{\mcitedefaultendpunct}{\mcitedefaultseppunct}\relax
\EndOfBibitem
\bibitem[ZHOU \latin{et~al.}(2010)ZHOU, KUANG, CHU, and YU]{ZHOUZ.Y2010Dpso}
ZHOU,~Z.~Y.; KUANG,~S.~B.; CHU,~K.~W.; YU,~A.~B. Discrete particle simulation
  of particle–fluid flow: model formulations and their applicability.
  \emph{Journal of fluid mechanics} \textbf{2010}, \emph{661}, 482--510\relax
\mciteBstWouldAddEndPuncttrue
\mciteSetBstMidEndSepPunct{\mcitedefaultmidpunct}
{\mcitedefaultendpunct}{\mcitedefaultseppunct}\relax
\EndOfBibitem
\bibitem[Olshanskii \latin{et~al.}(2009)Olshanskii, Lube, Heister, and
  Löwe]{OLSHANSKII20093975}
Olshanskii,~M.; Lube,~G.; Heister,~T.; Löwe,~J. Grad–div stabilization and
  subgrid pressure models for the incompressible Navier–Stokes equations.
  \emph{Computer Methods in Applied Mechanics and Engineering} \textbf{2009},
  \emph{198}, 3975--3988\relax
\mciteBstWouldAddEndPuncttrue
\mciteSetBstMidEndSepPunct{\mcitedefaultmidpunct}
{\mcitedefaultendpunct}{\mcitedefaultseppunct}\relax
\EndOfBibitem
\bibitem[Golshan \latin{et~al.}(2020)Golshan, Sotudeh-Gharebagh, Zarghami,
  Mostoufi, Blais, and Kuipers]{GolshanShahab2020Raio}
Golshan,~S.; Sotudeh-Gharebagh,~R.; Zarghami,~R.; Mostoufi,~N.; Blais,~B.;
  Kuipers,~J. Review and implementation of CFD-DEM applied to chemical process
  systems. \emph{Chemical engineering science} \textbf{2020}, \emph{221},
  115646--\relax
\mciteBstWouldAddEndPuncttrue
\mciteSetBstMidEndSepPunct{\mcitedefaultmidpunct}
{\mcitedefaultendpunct}{\mcitedefaultseppunct}\relax
\EndOfBibitem
\bibitem[Blais \latin{et~al.}(2019)Blais, Vidal, Bertrand, Patience, and
  Chaouki]{BlaisBruno2019EMiC}
Blais,~B.; Vidal,~D.; Bertrand,~F.; Patience,~G.~S.; Chaouki,~J. Experimental
  Methods in Chemical Engineering: Discrete Element Method—DEM.
  \emph{Canadian journal of chemical engineering} \textbf{2019}, \emph{97},
  1964--1973\relax
\mciteBstWouldAddEndPuncttrue
\mciteSetBstMidEndSepPunct{\mcitedefaultmidpunct}
{\mcitedefaultendpunct}{\mcitedefaultseppunct}\relax
\EndOfBibitem
\bibitem[Delacroix \latin{et~al.}(2020)Delacroix, Bouarab, Fradette, Bertrand,
  and Blais]{DELACROIX2020146}
Delacroix,~B.; Bouarab,~A.; Fradette,~L.; Bertrand,~F.; Blais,~B. Simulation of
  granular flow in a rotating frame of reference using the discrete element
  method. \emph{Powder Technology} \textbf{2020}, \emph{369}, 146--161\relax
\mciteBstWouldAddEndPuncttrue
\mciteSetBstMidEndSepPunct{\mcitedefaultmidpunct}
{\mcitedefaultendpunct}{\mcitedefaultseppunct}\relax
\EndOfBibitem
\bibitem[Gassmöller \latin{et~al.}(2018)Gassmöller, Lokavarapu, Heien,
  Puckett, and Bangerth]{GassmollerRene2018FaSP}
Gassmöller,~R.; Lokavarapu,~H.; Heien,~E.; Puckett,~E.~G.; Bangerth,~W.
  Flexible and Scalable Particle‐in‐Cell Methods With Adaptive Mesh
  Refinement for Geodynamic Computations. \emph{Geochemistry, geophysics,
  geosystems : G3} \textbf{2018}, \emph{19}, 3596--3604\relax
\mciteBstWouldAddEndPuncttrue
\mciteSetBstMidEndSepPunct{\mcitedefaultmidpunct}
{\mcitedefaultendpunct}{\mcitedefaultseppunct}\relax
\EndOfBibitem
\bibitem[Pepiot and Desjardins(2012)Pepiot, and Desjardins]{PEPIOT2012104}
Pepiot,~P.; Desjardins,~O. Numerical analysis of the dynamics of two- and
  three-dimensional fluidized bed reactors using an Euler–Lagrange approach.
  \emph{Powder Technology} \textbf{2012}, \emph{220}, 104--121, Selected Papers
  from the 2010 NETL Multiphase Flow Workshop\relax
\mciteBstWouldAddEndPuncttrue
\mciteSetBstMidEndSepPunct{\mcitedefaultmidpunct}
{\mcitedefaultendpunct}{\mcitedefaultseppunct}\relax
\EndOfBibitem
\bibitem[Elghobashi(1991)]{Elghobashi}
Elghobashi,~S. Particle-laden turbulent flows: Direct simulation and closure
  models. \emph{Applied Scientific Research} \textbf{1991}, \emph{48},
  301--314\relax
\mciteBstWouldAddEndPuncttrue
\mciteSetBstMidEndSepPunct{\mcitedefaultmidpunct}
{\mcitedefaultendpunct}{\mcitedefaultseppunct}\relax
\EndOfBibitem
\bibitem[Burstedde \latin{et~al.}(2011)Burstedde, Wilcox, and Ghattas]{p4est}
Burstedde,~C.; Wilcox,~L.~C.; Ghattas,~O. {\texttt{p4est}}: Scalable Algorithms
  for Parallel Adaptive Mesh Refinement on Forests of Octrees. \emph{SIAM
  Journal on Scientific Computing} \textbf{2011}, \emph{33}, 1103--1133\relax
\mciteBstWouldAddEndPuncttrue
\mciteSetBstMidEndSepPunct{\mcitedefaultmidpunct}
{\mcitedefaultendpunct}{\mcitedefaultseppunct}\relax
\EndOfBibitem
\bibitem[Dallavalle(1948)]{dallavalle}
Dallavalle,~J.~M. Micromeritics : the technology of fine particles. 1948\relax
\mciteBstWouldAddEndPuncttrue
\mciteSetBstMidEndSepPunct{\mcitedefaultmidpunct}
{\mcitedefaultendpunct}{\mcitedefaultseppunct}\relax
\EndOfBibitem
\bibitem[Di~Felice(1994)]{DiFeliceR1994Tvff}
Di~Felice,~R. The voidage function for fluid-particle interaction systems.
  \emph{International journal of multiphase flow} \textbf{1994}, \emph{20},
  153--159\relax
\mciteBstWouldAddEndPuncttrue
\mciteSetBstMidEndSepPunct{\mcitedefaultmidpunct}
{\mcitedefaultendpunct}{\mcitedefaultseppunct}\relax
\EndOfBibitem
\bibitem[Rong \latin{et~al.}(2013)Rong, Dong, and Yu]{RongL.W2013Lsof}
Rong,~L.; Dong,~K.; Yu,~A. Lattice-Boltzmann simulation of fluid flow through
  packed beds of uniform spheres: Effect of porosity. \emph{Chemical
  engineering science} \textbf{2013}, \emph{99}, 44--58\relax
\mciteBstWouldAddEndPuncttrue
\mciteSetBstMidEndSepPunct{\mcitedefaultmidpunct}
{\mcitedefaultendpunct}{\mcitedefaultseppunct}\relax
\EndOfBibitem
\bibitem[Jajcevic \latin{et~al.}(2013)Jajcevic, Siegmann, Radeke, and
  Khinast]{JAJCEVIC2013298}
Jajcevic,~D.; Siegmann,~E.; Radeke,~C.; Khinast,~J.~G. Large-scale CFD–DEM
  simulations of fluidized granular systems. \emph{Chemical Engineering
  Science} \textbf{2013}, \emph{98}, 298--310\relax
\mciteBstWouldAddEndPuncttrue
\mciteSetBstMidEndSepPunct{\mcitedefaultmidpunct}
{\mcitedefaultendpunct}{\mcitedefaultseppunct}\relax
\EndOfBibitem
\bibitem[Beetstra \latin{et~al.}(2007)Beetstra, van~der Hoef, and
  Kuipers]{beetstradrag}
Beetstra,~R.; van~der Hoef,~M.~A.; Kuipers,~J. A.~M. Drag force of intermediate
  Reynolds number flow past mono- and bidisperse arrays of spheres. \emph{AIChE
  Journal} \textbf{2007}, \emph{53}, 489--501\relax
\mciteBstWouldAddEndPuncttrue
\mciteSetBstMidEndSepPunct{\mcitedefaultmidpunct}
{\mcitedefaultendpunct}{\mcitedefaultseppunct}\relax
\EndOfBibitem
\bibitem[Norouzi \latin{et~al.}(2021)Norouzi, Golshan, and
  Zarghami]{NorouziHamidReza2021Otdf}
Norouzi,~H.~R.; Golshan,~S.; Zarghami,~R. On the drag force closures for
  multiphase flow modeling. \emph{Chemical product and process modeling}
  \textbf{2021}, \relax
\mciteBstWouldAddEndPunctfalse
\mciteSetBstMidEndSepPunct{\mcitedefaultmidpunct}
{}{\mcitedefaultseppunct}\relax
\EndOfBibitem
\bibitem[Bérard \latin{et~al.}(2020)Bérard, Patience, and
  Blais]{BerardAriane2020Emic}
Bérard,~A.; Patience,~G.~S.; Blais,~B. Experimental methods in chemical
  engineering: Unresolved CFD‐DEM. \emph{Canadian journal of chemical
  engineering} \textbf{2020}, \emph{98}, 424--440\relax
\mciteBstWouldAddEndPuncttrue
\mciteSetBstMidEndSepPunct{\mcitedefaultmidpunct}
{\mcitedefaultendpunct}{\mcitedefaultseppunct}\relax
\EndOfBibitem
\bibitem[ERGUN(1952)]{Ergun}
ERGUN,~S. Fluid flow through packed columns. \emph{Chem. Eng. Prog.}
  \textbf{1952}, \emph{48}, 89--94\relax
\mciteBstWouldAddEndPuncttrue
\mciteSetBstMidEndSepPunct{\mcitedefaultmidpunct}
{\mcitedefaultendpunct}{\mcitedefaultseppunct}\relax
\EndOfBibitem
\bibitem[Wen and Yu(1966)Wen, and Yu]{wen1966generalized}
Wen,~C.; Yu,~Y. A generalized method for predicting the minimum fluidization
  velocity. \emph{AIChE Journal} \textbf{1966}, \emph{12}, 610--612\relax
\mciteBstWouldAddEndPuncttrue
\mciteSetBstMidEndSepPunct{\mcitedefaultmidpunct}
{\mcitedefaultendpunct}{\mcitedefaultseppunct}\relax
\EndOfBibitem
\bibitem[Noda \latin{et~al.}(1986)Noda, Uchida, Makino, and Kamo]{NODA1986149}
Noda,~K.; Uchida,~S.; Makino,~T.; Kamo,~H. Minimum fluidization velocity of
  binary mixture of particles with large size ratio. \emph{Powder Technology}
  \textbf{1986}, \emph{46}, 149--154\relax
\mciteBstWouldAddEndPuncttrue
\mciteSetBstMidEndSepPunct{\mcitedefaultmidpunct}
{\mcitedefaultendpunct}{\mcitedefaultseppunct}\relax
\EndOfBibitem
\bibitem[Anantharaman \latin{et~al.}(2018)Anantharaman, Cocco, and
  Chew]{ANANTHARAMAN2018454}
Anantharaman,~A.; Cocco,~R.~A.; Chew,~J.~W. Evaluation of correlations for
  minimum fluidization velocity (Umf) in gas-solid fluidization. \emph{Powder
  Technology} \textbf{2018}, \emph{323}, 454--485\relax
\mciteBstWouldAddEndPuncttrue
\mciteSetBstMidEndSepPunct{\mcitedefaultmidpunct}
{\mcitedefaultendpunct}{\mcitedefaultseppunct}\relax
\EndOfBibitem
\bibitem[Snider(2001)]{SniderD.M2001AITM}
Snider,~D. An Incompressible Three-Dimensional Multiphase Particle-in-Cell
  Model for Dense Particle Flows. \emph{Journal of computational physics}
  \textbf{2001}, \emph{170}, 523--549\relax
\mciteBstWouldAddEndPuncttrue
\mciteSetBstMidEndSepPunct{\mcitedefaultmidpunct}
{\mcitedefaultendpunct}{\mcitedefaultseppunct}\relax
\EndOfBibitem
\bibitem[Dimonte \latin{et~al.}(2004)Dimonte, Youngs, Dimits, Weber, Marinak,
  Wunsch, Garasi, Robinson, Andrews, Ramaprabhu, Calder, Fryxell, Biello,
  Dursi, MacNeice, Olson, Ricker, Rosner, Timmes, Tufo, Young, and
  Zingale]{dimonte}
Dimonte,~G. \latin{et~al.}  A comparative study of the turbulent
  Rayleigh–Taylor instability using high-resolution three-dimensional
  numerical simulations: The Alpha-Group collaboration. \emph{Physics of
  Fluids} \textbf{2004}, \emph{16}, 1668--1693\relax
\mciteBstWouldAddEndPuncttrue
\mciteSetBstMidEndSepPunct{\mcitedefaultmidpunct}
{\mcitedefaultendpunct}{\mcitedefaultseppunct}\relax
\EndOfBibitem
\bibitem[Yue \latin{et~al.}(2020)Yue, Wang, Bahl, de~Silva, and
  Shen]{YueYuanhe2020EIoS}
Yue,~Y.; Wang,~S.; Bahl,~P.; de~Silva,~C.; Shen,~Y. Experimental Investigation
  of Spout Deflection in a Rectangular Spouted Bed by the PIV Method.
  \emph{Industrial \& engineering chemistry research} \textbf{2020}, \emph{59},
  13810--13819\relax
\mciteBstWouldAddEndPuncttrue
\mciteSetBstMidEndSepPunct{\mcitedefaultmidpunct}
{\mcitedefaultendpunct}{\mcitedefaultseppunct}\relax
\EndOfBibitem
\bibitem[Arndt \latin{et~al.}(2016)Arndt, Braack, and
  Lube]{ArndtDaniel2016FEft}
Arndt,~D.; Braack,~M.; Lube,~G. \emph{Numerical Mathematics and Advanced
  Applications ENUMATH 2015}; Lecture Notes in Computational Science and
  Engineering; Springer International Publishing: Cham, 2016; pp 95--103\relax
\mciteBstWouldAddEndPuncttrue
\mciteSetBstMidEndSepPunct{\mcitedefaultmidpunct}
{\mcitedefaultendpunct}{\mcitedefaultseppunct}\relax
\EndOfBibitem
\end{mcitethebibliography}

\end{document}